\documentclass[12pt]{article}

\usepackage{amsmath}
\usepackage{mathrsfs}
\usepackage{xcolor}

\usepackage{draft} 
\usepackage{hyperref}
\usepackage{graphicx,color,subfig}
\usepackage{cite}
\usepackage{mciteplus}
\usepackage{skak}
\usepackage{empheq}
\usepackage{tikz}
\usepackage{booktabs}
\usepackage{siunitx}
\usepackage{bbm}
\usepackage{makecell}
\usepackage{float}

\usepackage{bm}
\usepackage{braket}
\usepackage{slashed}

\DeclareFontFamily{OT1}{pzc}{}
\DeclareFontShape{OT1}{pzc}{m}{it}{<-> s * [1.10] pzcmi7t}{}
\DeclareMathAlphabet{\mathpzc}{OT1}{pzc}{m}{it}

\def\be#1\ee{\begin{align}#1\end{align}}

\makeatletter
\newcommand{\bdryno}{\mathpalette\bdry@no\relax}
\newcommand{\bdry@no}[2]{%
  \mspace{1mu}%
  \vbox{%
    \hbox{$\m@th#1\scriptstyle{\ast}$}
    \nointerlineskip
    \kern.25ex
    \hbox{$\m@th#1\scriptstyle{\ast}$}
    \kern-.06ex
  }%
  \mspace{1mu}%
}
\makeatother

\newcommand*\dif{\mathop{}\!\mathrm{d}}

\newenvironment{fix}{\color{red}}{\ignorespacesafterend}

\begin{document}

\unitlength = .8mm

\begin{titlepage}

\begin{center}

\hfill \\
\hfill \\
\vskip 1cm
\title{On the Equivalence between SRS and PCO Formulations of Superstring Perturbation Theory}

\author{Charles Wang and Xi Yin}

\address{
Jefferson Physical Laboratory, Harvard University, \\
Cambridge, MA 02138 USA
}

\email{charles\_wang@g.harvard.edu, xiyin@fas.harvard.edu}

\abstract{
We establish the equivalence between two formulations of superstring perturbation theory, one based on integration over the supermoduli space of super Riemann surfaces (SRS), the other based on integration over the bosonic moduli space with insertions of picture changing operators (PCO) on the worldsheet and the vertical integration prescription, by showing how the latter arises from a specific construction of the supermoduli integration contour.
}

\vfill

\end{center}

\end{titlepage}

\eject

\begingroup
\hypersetup{linkcolor=black}

\tableofcontents

\endgroup

\section{Introduction}

A cornerstone of superstring theory is the formulation of perturbative closed string scattering amplitudes based on the path integral of two-dimensional supergravity coupling to matter fields on the worldsheet. The precise definition of the worldsheet path integral requires fixing the gauge redundancy of super-diffeomorphism and super-Weyl invariance. There are two standard gauge choices. The first one sets the worldsheet gravitino field to zero on local coordinate patches. The gauge choices for two different patches generally differ on the overlap by a superconformal transformation, giving rise to a super-Riemann surface (SRS) structure of the worldsheet \cite{DHoker:1988pdl, Witten:2012ga}. In this ``SRS formalism", the asymptotic closed string states are represented as punctures, of Neveu-Schwarz (NS) or Ramond (R) type, and their S-matrix elements are formulated as integrals of a form $\Omega$, defined through correlation functions of a superconformal field theory (SCFT) on the worldsheet SRS, over the supermoduli space of the latter. More precisely, for type II superstring considered in this paper, the integration of $\Omega$ is defined over a super contour $\mathfrak{S}$ of $\mathfrak{M}\times\overline{\mathfrak M}$, where $\mathfrak{M}$ and $\overline{\mathfrak M}$ are the supermoduli spaces of a pair of holomorphic and anti-holomorphic SRSs.

A second gauge choice is to set the worldsheet gravitino to a distribution supported at a finite set of points on the worldsheet. Up to regularization ambiguities that can be fixed by requirement of BRST invariance at the quantum level, integration over the gauge-inequivalent degrees of freedom of the gravitino gives rise to the insertions of picture-changing operators (PCOs) on the worldsheet equipped with an ordinary (bosonic) Riemann surface structure \cite{Friedan:1985ge, Verlinde:1987sd}. The space of states or vertex operators of the worldsheet SCFT is graded according to the picture number, of integer and half-integer values in the NS and R sectors respectively. The asymptotic string states are represented by BRST-closed vertex operators subject to Siegel constraint, in the NS sector of picture number $-1$, or in the R sector of picture number $-{1\over 2}$. The data of the worldsheet geometry together with the locations of PCOs are parameterized by a fiber bundle ${\cal Y}$ over the moduli space ${\cal M}$ of ordinary Riemann surfaces. The spacetime S-matrix elements are formulated as integrals of a form $\Omega$, defined through worldsheet SCFT correlators with PCO insertions, over a suitable contour ${\cal S}\subset{\cal Y}$. We refer to this as the PCO formalism, and will refer to ${\cal S}$ as the ``PCO contour".

Heuristically, the SRS and PCO formalisms of superstring amplitudes come from different gauge choices of the worldsheet path integral, and are anticipated to be equivalent by virtue of the Faddeev-Popov/BRST quantization procedure. However, as the path integral is only well-defined after gauge fixing, to establish such an equivalence requires demonstrating that the two formalisms are related by BRST-exact deformations at the quantum level, which is not at all obvious. Furthermore, the gauge choice that leads to PCOs may suffer from ``spurious singularities" that occur along complex codimension 1 loci in the fiber bundle ${\cal Y}$ \cite{Verlinde:1987sd, Atick:1987rk}, an issue that cast doubt on the consistency of the PCO formalism for over two decades. 

The problem of spurious singularity was overcome in \cite{Sen:2014pia, Sen:2015hia} through the prescription of ``vertical integration". The basic idea is that the PCO contour ${\cal S}$ need not be a section of the bundle ${\cal Y}\to {\cal M}$, nor even a cycle; it suffices for ${\cal S}$ to be a chain that projects onto ${\cal M}$, such that the boundary of ${\cal S}$ consists of only ``vertical" components that extend in the fiber directions, on which a PCO moves along a closed path on a given Riemann surface. This allows for ${\cal S}$ to evade the loci of spurious singularities, while preserving the BRST-invariance and unitarity of superstring amplitudes.

The SRS formalism involves integrations over supermanifolds that, while in principle defined say via partition of unity, are often difficult to evaluate due to lack of convenient parameterizations of the supermoduli space \cite{Witten:2012bg, Donagi:2013dua}. In contrast, the PCO formalism gives a relatively straightforward algorithm for evaluating superstring amplitudes (numerically if necessary), and has been further extended to define off-shell amplitudes and vertices of superstring field theory \cite{Sen:2014pia, deLacroix:2017lif}. On the other hand, the choice of the PCO contour ${\cal S}$ is highly non-canonical. Nonetheless, it has been shown in \cite{Sen:2014pia, Sen:2015hia, Erler:2017dgr} that deformations of ${\cal S}$ do not affect on-shell amplitudes.

In this paper, we establish the equivalence between the two formalisms by showing that a specific choice of the super contour $\mathfrak{S}$, upon patchwise integration over fermionic coordinates, produces the PCO formalism. Our construction of $\mathfrak{S}$ consists of ``horizontal patches" in the supermoduli space that corresponds to PCOs at fixed locations, and interpolating ``vertical patches" that amount to vertical integration. In defining a patch of $\mathfrak{S}$ as the image of a map $\mathfrak{I}$ into the supermoduli space, and the integration of $\Omega$ via that of the pullback $\mathfrak{I}^*\Omega$, it is important to ensure that $\mathfrak{I}$ has non-singular Berezinian \cite{Witten:2012bg}. The latter is closely related to the evasion of spurious singularity in the PCO formalism.

In section \ref{sec:srs}, we review the notion of SRS, its supermoduli space, the meaning of a SCFT on a general SRS, and the formulation of type II superstring amplitudes as an integral over the supermoduli space. Section \ref{sec:pco} reviews the PCO, spurious singularities, and the integration contour used to define superstring amplitudes in the PCO formalism. In section \ref{sec:integration}, we explain the definition of integral over a supermanifold $\mathfrak{M}$ via partition of unity, and then introduce an alternative way of performing the integration by ``lifting and interpolation". The basic idea is that one can cover $\mathfrak{M}$ with super coordinate patches $\mathfrak{U}_\A$, each of which projects onto a bosonic patch ${\cal U}_\A$ of reduced space ${\cal M}$ via the projection map $\pi_\A: \mathfrak{U}_\A\to {\cal U}_\A$. Dividing ${\cal M}$ into cells ${\cal D}_\A\subset{\cal U}_\A$, one can reduce the integration over $\pi_\A^{-1}({\cal D}_\A)$ to a bosonic one by integrating out the fermionic fibers. However, the fermionic fibers of $\pi_\A$ and $\pi_\B$ generally do not agree on the boundary between adjacent cells, ${\cal D}_\A\cap {\cal D}_\B$. This is corrected for by an interpolating chain between $\pi_\A^{-1}({\cal D}_\A)$ and $\pi_\B^{-1}({\cal D}_\B)$ constructed as the image of an interpolation map $\mathfrak{I}_{\A\B}$. To recover the full integral over $\mathfrak{M}$, one must include the interpolating chains that account for the mismatch of fermionic fibers along all-codimensional interfaces between adjacent cells. An explicit construction of the super coordinate patches $\mathfrak{U}_\A$ of the supermoduli space $\mathfrak{M}$, and the transition maps between them, is given in section \ref{sec:lift}. The general structure of the interpolating chains for the supermoduli integration contour $\mathfrak{S}\subset \mathfrak{M}\times\overline{\mathfrak M}$ of type II superstring theory is described in section \ref{sec:supercontour}.

Section \ref{sec:mtos} is the core of the paper. In section \ref{sec:pcoemerge}, we explain how PCO insertions arise from the integration over the fiber of the projection map $\pi_\A$ of section \ref{sec:lift}. The interpolation map ${\mathfrak I}_{\A\B}$ between the fibers of adjacent patches is constructed in section \ref{sec:vertemerge}, and shown to reproduce the vertical integration in the codimension 1 case. This is then generalized to the higher codimension case in section \ref{sec:pcohighercodim}. An application of this general construction in the example of genus two SRSs will be discussed in a follow-up paper \cite{paper:g2}. Some concluding remarks are given in section \ref{sec:discuss}.

\section{The SRS formalism}
\label{sec:srs}

\subsection{SRS and its moduli space}

A super-Riemann surface (SRS) ${\mathfrak C}$ is a $1|1$ complex supermanifold, equipped with a totally non-integrable rank $0|1$ sub-bundle of its tangent bundle. It can be covered with super coordinate charts $U_i$, parameterized by bosonic and fermionic coordinates $(z_i,\theta_i)$, such that the transition map on the overlap $U_i\cap U_j$ takes form of a superconformal transformation
\ie\label{sctrans}
&z_i = f_{ij}(z_j) + \theta_j g_{ij}(z_j) h_{ij}(z_j),
\\
& \theta_i = g_{ij}(z_j) + \theta_j h_{ij}(z_j),~~~ {\rm with}~ h_{ij} = \pm\sqrt{\partial f_{ij} + g_{ij}\partial g_{ij}}.
\fe
Generally, the transition functions $f_{ij}$ and $g_{ij}$ can be deformed by a set of bosonic and fermionic parameters, while each $h_{ij}$ involves a choice of sign. Modulo superconformal diffeomorphism, the former give rise to the even and odd moduli, denoted by $\tau^m$ ($m=1,\cdots, d_{ e}$) and $\nu^a$ ($a=1,\cdots, d_{ o}$) respectively, whereas the latter give rise to a choice of spin structure, denoted by $\epsilon$. We will sometimes indicate the moduli dependence of the SRS with the notation $\mathfrak{C}_{t,\nu}$, or simply $\mathfrak{C}_\nu$ to indicate the odd moduli dependence.

By setting all the odd parameters in the transition maps that define $\mathfrak{C}$ to zero, which includes setting all of $g_{ij}$'s to zero, we obtain the reduced space $\Sigma$ which is an ordinary Riemann surface. Conversely, given an ordinary Riemann surface $\Sigma$ with a choice of spin structure $\epsilon$, there is a corresponding SRS $\mathfrak{C}_0$ with $g_{ij}= 0$, known as a split SRS.

For our purpose it is important to extend the definition of SRS to allow for punctures of NS and R type. The NS puncture amounts to marking a point, say $(z,\theta)=(0,0)$, on a super disc $D$ with coordinates $(z,\theta)$. The R puncture is defined by further splitting the disc $D$ into two wedges, parameterized by $(z,\theta)$ and $(z',\theta')$ such that the transition map between them takes the form $z'=z$, $\theta'=\pm\theta$, where the sign is the opposite on the two disjoint wedges of the overlap.

The string worldsheet in the superconformal gauge, where the worldsheet gravitino are set to zero patch wise, can be viewed as a diagonal subspace of $\mathfrak{C}\times \overline{\mathfrak{C}}'$, where $\overline{\mathfrak{C}}'$ is a SRS whose reduced space is $\overline\Sigma$, the complex conjugate Riemann surface of $\Sigma$. Note that the choice of spin structure of $\overline{\mathfrak{C}}'$ and the types of its punctures are a priori independent from those of $\mathfrak{C}$.

The SRS $\mathfrak{C}$ can be classified topologically according to the genus $g$ of its reduced space, the number of NS and R punctures, $n_{\rm NS}$ and $n_{\rm R}$, and the spin structure $\epsilon$. The supermoduli space of such SRS, denoted $\mathfrak{M}_{g, n_{\rm NS}, n_{\rm R}, \epsilon}$ or simply $\mathfrak{M}$, is a complex supermanifold of dimension $d_{ e}|d_{ o}$, where $d_{ e}=3g-3+n_{\rm NS} + n_{\rm R}$, $d_o = 2g-2 + n_{\rm NS} + {1\over 2}n_{\rm R}$. Its reduced space ${\cal M}_{g, n_{\rm NS}, n_{\rm R},\epsilon}$ is the moduli space of an ordinary Riemann surface of genus $g$, with $n_{\rm NS}+n_{\rm R}$ punctures and a choice of spin structure.

One should be cautious that a supermanifold is defined not as a set but as a locally ringed space. In particular, the operation of ``forgetting the fermionic coordinates" does not define a map from $\mathfrak{M}$ to its reduced space ${\cal M}$, as the pullback of a function under such an operation does not give a well-defined function on the supermanifold $\mathfrak{M}$. For our purpose, it is important to have a notion of integration of super forms on a supermanifold, that may be defined via partition of unity and respects Stokes' theorem, which we review in section \ref{sec:integration} (see also \cite{Witten:2012bg, Witten:2012ga}).

\subsection{SCFT on SRS }
\label{sec:srsscft}

Throughout this paper we will consider a superconformal field theory(SCFT) of central charge $c=0$ on the worldsheet, with holomorphic stress-energy tensor $T(z)$ and supercurrent $G(z)$ that obey the super-Virasoro algebra, as well as their anti-holomorphic counter parts. In particular, the singular part of the OPE of a pair of supercurrents takes the form
\ie\label{ggope}
G(z) G(0) \sim {2\over z} T(0).
\fe
With the absence of Weyl anomaly, the SCFT is a priori defined on a Riemann surface $\Sigma$ equipped with a spin structure $\epsilon$ that specifies the (anti-)periodicity of the supercurrent via transport along closed paths. Equivalently, we may view this as defining the SCFT on the corresponding split SRS $\mathfrak{C}_0$.

This notion of SCFT can be extended to that over an arbitrary SRS by deforming away from the split case, as follows. Given a set of super coordinate charts $U_i$ of $\mathfrak{C}$, parameterized by $(z_i, \theta_i)$, a deformation of the supermoduli amounts to a deformation of the transition maps (\ref{sctrans}). This is represented at the level of SCFT correlators by inserting 
\ie\label{scdeform}
\delta{\cal T} = - \sum_{(ij)} \int_{C_{ij}} {[dz_i|  d\theta_i ]\over 2\pi i}\, \mathbb{T}(z_i, \theta_i)\left. \big[ \delta z_i - (\delta \theta_i) \theta_i \big]\right|_{z_j,\theta_j},
\fe
where $\mathbb{T}(z,\theta) \equiv {1\over 2} G(z) + \theta T(z)$ is the super stress tensor, and $C_{ij}$ are a set of arcs defined as follows. Let $D_i$ be a closed domain contained within the bosonic part of $U_i$, such that $D_i$ and $D_j$ only meet along their boundary, and the union of $D_i$'s is the entirety of $\Sigma$. $C_{ij}$ is the arc along $D_i\cap D_j$, oriented such that $\sum_j C_{ij} = \partial D_i$. The superconformal covariance of (\ref{scdeform}) ensures that two different sets of coordinate charts and transition maps describe equivalent SRS if the corresponding super stress tensor insertions (\ref{scdeform}) are equivalent (in the SCFT sense) via contour deformations and OPEs as governed by the superconformal algebra. 

A particularly useful parameterization of the odd moduli is through the gluing map of super discs, as follows. Starting with a split SRS $\mathfrak{C}_0$ with its reduced space $\Sigma$, we choose a set of points $z_a$ on $\Sigma$, $a=1,\cdots, d_o$, along with sufficient small discs $D_a$ centered at $z_a$. Denote by $U_a$ a coordinate patch of $\Sigma$ that contains $D_a$, which is further split into two patches $U_a = U_a'\cup D_a$, where $U_a'$ is an annulus that does not contain $z_a$. Each of these patches is promoted to a super coordinate chart of $\mathfrak{C}_0$, which we will denote by the same symbol. Let $(z,\theta)$ be the coordinates on the super chart of $U_a'$, and $(w,\eta)$ be the coordinates on the super chart of $D_a$, such that the transition map on their overlap in $\mathfrak{C}_0$ is simply $(w,\eta)=(z,\theta)$. Now we can construct a new SRS $\mathfrak{C}_\nu$ by deforming these transition maps to
\ie\label{superdiscglue}
& w = z - {\theta\nu^a\over z-z_a},
\\
& \eta = \theta - {\nu^a\over z-z_a},
\fe
where $\nu^a$ is a fermionic parameter, for each $a=1,\cdots,d_o$. In the language of SCFT, the deformation from $\mathfrak{C}_0$ to $\mathfrak{C}_\nu$ amounts to the insertion of
\ie\label{nvgdef}
\prod_a \left[ 1 - \oint_{\partial D_a} {dw\over 2\pi i} G(w) {\nu^a \over w-z_a} \right] = \prod_a \left[1+ \nu^a G(z_a) \right]
\fe
on $\Sigma$. 

For generic choice of $z_1,\cdots, z_{d_o}$, the $\nu^a$'s give a set of non-degenerate local fermionic coordinates on the supermoduli space. Degeneration occurs if there is a weight $-{1\over 2}$ meromorphic differential $r(z) (dz)^{-{1\over 2}}$ with only $d_o$ simple poles at the $z_a$ with residue $r_a$, so that the linear combination of supercurrent insertions $\sum_a r_a G(z_a) = \oint_{\sqcup \partial D_a} {dz\over 2\pi i} r(z) G(z)$ vanishes by shrinking the contour $\sqcup_a \partial D_a$ away from $\{z_a\}$. This is precisely the condition for a spurious singularity (see section \ref{sec:pcointro}). 


\subsection{Superstring amplitude}
\label{sec:superamp}

In type II superstring theory, the perturbative scattering amplitude of $n$ closed string asymptotic states, represented by BRST-closed vertex operators ${\cal V}_i$ that obey Siegel constraint, takes the general form
\ie
{\cal A}[{\cal V}_1,\cdots,{\cal V}_n] = \sum_{g\geq 0} {g_s^{2g-2}\over 2^{2g}} \sum_{\epsilon, \widetilde\epsilon}  {\cal A}_{g,\epsilon, \widetilde\epsilon}[{\cal V}_1,\cdots,{\cal V}_n],
\fe
where $g$ is the genus and $\epsilon,\widetilde\epsilon$ label the holomorphic and anti-holomorphic spin structures, respectively. For simplicity of exposition we will focus on the case where all of the asymptotic string states are of (NS, NS) type. The distinction between type IIA and type IIB strings is reflected in the choice of overall sign of the odd spin structure contributions to the amplitude. We have
\ie\label{supermoduliintegral}
&{\cal A}_{g,\epsilon,\widetilde\epsilon}[{\cal V}_1,\cdots,{\cal V}_n] = \int_{\mathfrak{S}} \Omega,
\fe
where the integration contour $\mathfrak{S} \subset \mathfrak{M}_{g,n,\epsilon} \times \overline{{\mathfrak M}_{g,n,\widetilde\epsilon}}$ is a subspace of codimension $d_e|0$, whose reduced space is the diagonal subspace of ${\cal M}_{g,n,\epsilon} \times \overline{{\cal M}_{g,n,\widetilde\epsilon}}$. $\Omega$ is the super integral  form defined in a neighborhood of $\mathfrak{S}$ in $\mathfrak{M}_{g,n,\epsilon} \times \overline{{\mathfrak M}_{g,n,\widetilde\epsilon}}$, given by the following correlator of the worldsheet SCFT on the SRS $\mathfrak{C}\times \overline{\mathfrak{C}'}$,
\ie\label{omegaintg}
& \Omega = \left\langle e^{\mathfrak{B}} \prod_{i=1}^n {\cal V}_i \right\rangle,
\fe
where $\mathfrak{B}$ is the 1-form
\ie
\mathfrak{B} =\sum_{k=1}^{2d_e} {\cal B}_{t^k} dt^k + \sum_{a=1}^{d_o} \delta(d\nu^a)  \delta({\cal  B}_{\nu^a}) + c.c.,
\fe
with
\ie{}
& {\cal B}_{t^k} =  \sum_{(ij)} \int_{C_{ij}} {[dz_i|d\theta_i] \over 2\pi i} \, \mathbb{B}(z_i, \theta_i)\left. \left[ {\partial z_i\over \partial t^k} - {\partial \theta_i\over \partial t^k} \theta_i \right]\right|_{z_j,\theta_j},
\\
& {\cal B}_{\nu^a} =  \sum_{(ij)} \int_{C_{ij}} {[dz_i|d\theta_i] \over 2\pi i} \, \mathbb{B}(z_i, \theta_i)\left. \left[ {\partial z_i\over \partial\nu^a} - {\partial \theta_i\over \partial\nu^a} \theta_i \right]\right|_{z_j,\theta_j} .
\fe
Here $\mathbb{B}=\B(z) + \theta b(z)$. The domain of integration is defined as in (\ref{scdeform}). Here $t^k=(\tau^m, \bar\tau^m)$ are bosonic moduli parameters. We emphasize that the superconformal transition maps need not depend holomorphically on $\tau^m$. We have also adopted the convention of \cite{Witten:2012bg} in which the 1-form $dt^k$ is Grassmann-odd, $d\nu^a$ is Grassmann-even, and the integral form $\delta(d\nu^a)$ is Grassmann-odd. 

Note that ${\cal B}_{\nu^a}$ is a Grassmann-even object, and $\delta({\cal B}_{\nu^a})$ is generally not a local operator. The latter should be understood as a holomorphic distribution that serves to absorb the integration over zero modes of $\B$. Typically, ${\cal B}_{\nu^a}$ takes the form 
\ie
{\cal B}_{\nu^a} = \int_C {dz\over 2\pi i} \B(z) g_a(z) +\cdots,
\fe
where $g_a(z)$ is a holomorphic vector field defined in a neighborhood of the path $C$, and $\cdots$ represents terms that involve odd moduli. The functional integration over $\B$ involves an integration over zero modes of the form $\B(z) =\sum_k \B_{(k)} \lambda^{(k)}(z)$, where $\lambda^{(k)}(z) (dz)^{3\over 2}$ are a basis of meromorphic weight ${3\over 2}$ forms that are compatible with the spin structure $\epsilon$ and have simple poles at the location of vertex operators ${\cal V}_1,\cdots,{\cal V}_n$. The result of the zero mode integral with $\delta({\cal B}_{\nu^a})$ insertion is
\ie
\int \prod_k d\B^{(k)} \prod_a \delta({\cal B}_{\nu^a}) = {1\over \det M_a^k},~~~~M_a^k = \int_C {dz\over 2\pi i} \lambda^{(k)}(z) g_a(z).
\fe
The matrix $M_a^k$ is non-degenerate provided that $\nu^a$ are non-degenerate fermionic coordinates on the supermoduli space. Consequently, the integrand $\Omega$ is non-singular at least away from the boundary of the moduli space.

\section{The PCO formalism}
\label{sec:pco}

\subsection{The PCO and spurious singularities}
\label{sec:pcointro}

Gauge fixing the worldsheet supergravity path integral by setting the gravitino to a sum over delta functions \cite{Verlinde:1987sd, DHoker:1988pdl}, upon suitable regularization, leads to the insertion of PCO, defined by
\ie\label{pcodef}
{\cal X}(z) &= - Q_B \cdot\Theta(\B(z)) = {1\over 2} \lim_{w\to z} \left[ G(w) \delta(\B(z)) - {1\over w-z} b(z) \delta'(\B(z)) \right] - {1\over 4} \partial b(z) \delta'(\B(z))
\fe
and similarly for its anti-holomorphic counter part $\widetilde{\cal X}(\bar z)$. 

It is convenient to pass to an equivalent representation of the $\B\C$ system in terms of a linear dilaton field $\phi$ and Grassmann-odd fields $(\xi,\eta)$\footnote{The nontrivial OPEs among $\phi,\xi,\eta$ are $\phi(z)\phi(0)\sim -\ln z$, $\xi(z) \eta(0)\sim {1\over z}$.} \cite{Friedan:1985ge}, with $\B\simeq e^{-\phi}\partial\xi$, $\C\simeq \eta e^\phi$, $\delta(\B)\simeq e^\phi$, $\delta(\C)\simeq e^{-\phi}$. We adopt a cocycle convention in which $e^{\A\phi}$ is Grassmann odd for odd integer $\A$. In the $(\phi,\xi,\eta)$ representation, the PCO is written as
\ie
{\cal X}(z) &= Q_B \cdot \xi(z) = - {1\over 2} e^\phi G^{\rm matter} + c\partial \xi - {1\over 4} e^{2\phi} \partial\eta b - {1\over 4} \partial( e^{2\phi} \eta b).
\fe

The correlators of an arbitrary set of $\xi$, $\eta$, and $e^{\A\phi}$ insertions on any Riemann surface $\Sigma$ with a choice of spin structure are given in terms of theta functions on the Jacobian variety of $\Sigma$ and prime forms in \cite{Verlinde:1987sd}. An important feature of such correlators is the presence of so-called spurious singularities, which occur whenever there is a zero mode of $\C(z)$, that is, a weight $-{1\over 2}$ meromorphic differential with a simple pole at every $\delta(\B)$ insertion (or the location of a PCO) and a simple zero at every $\delta(\C)$ insertion (or location of a vertex operator for an NS string state). 

Let us remark that the correlators of $\delta(\B)$ and $\delta(\C)$ can be used to construct half-integer weight meromorphic differentials of certain prescribed zeros and poles, which will be used extensively later. For instance, given a generic set of $n$ points $\{w_i\}$ and $2g-1+n$ points $\{z_a\}$ on $\Sigma$, there is a weight $-{1\over 2}$ meromorphic differential $\zeta = f(z) dz$ with simple poles at $z_a$ and zeros at $w_i$, that is unique up to constant rescaling, given by
\ie\label{meroweighthalf}
f(z) = {1\over \left\langle \delta(\C(z)) \prod_{i=1}^n \delta(\C(w_i)) \prod_{a=1}^{2g-1+n} \delta(\B(z_a)) \right\rangle_{\Sigma,\epsilon}}.
\fe

\subsection{The integration contour}

We will refer to the diagonal subspace ${\cal M}\subset {\cal M}_{g,n,\epsilon}\times \overline{{\cal M}_{g,n,\widetilde\epsilon}}$ (i.e. the reduced space of the supermanifold $\mathfrak{S}$ of section \ref{sec:superamp}) as the bosonic moduli space of the Riemann surface $\Sigma$ with holomorphic spin structure $\epsilon$ and anti-holomorphic spin structure $\widetilde\epsilon$. Let $\pi:{\cal Y}\to {\cal M}$ be the fiber bundle whose fiber 
\ie
\pi^{-1}(x)\simeq (\Sigma_x\times \overline\Sigma_x)^{d_o}
\fe
is the space of the loci of $d_o$ pairs of holomorphic and anti-holomorphic PCOs inserted on the surface $\Sigma_x$ corresponding to the point $x\in {\cal M}$.

In the formalism of type II string amplitude based on PCOs \cite{Sen:2014pia, Sen:2015hia}, the genus $g$ amplitude takes the form
\ie\label{pcoamp}
{\cal A}_{g,\epsilon,\widetilde\epsilon}[{\cal V}_1,\cdots,{\cal V}_n] = \int_{\cal{S}} \widetilde\Omega.
\fe
Here $\widetilde\Omega$ is a  form on ${\cal Y}$ defined as the worldsheet SCFT correlator
\ie\label{tildeomg}
\widetilde\Omega = \left\langle e^{\pi^*{\cal B}}  \prod_{i=1}^n {\cal V}_i \prod_{a=1}^{d_o} \big[ {\cal X}(z_a)+d\xi(z_a)\big]  \big[ \widetilde{\cal X}(\bar z_a)+d\widetilde\xi(\bar z_a) \big] \right\rangle_{\Sigma,\epsilon,\widetilde\epsilon},
\fe
where $(z_a,\bar z_a)$ are viewed as coordinates on the fiber of ${\cal Y}\to {\cal M}$. ${\cal B}$ is the 1-form on ${\cal M}$ defined by
\ie
{\cal B} =  \sum_{(ij)} \int_{C_{ij}} {dz_i \over 2\pi i} \, b(z_i)\left.  {\partial z_i\over \partial t^k}\right|_{z_j,\theta_j} dt^k + c.c.
\fe
The integration contour $\cal{S}$ in (\ref{pcoamp}) is a $d_e$-dimensional chain in ${\cal Y}$ that evades spurious singularities of $\widetilde\Omega$, such that $\pi(\cal{S})$ covers ${\cal M}$ once. Importantly, $\cal{S}$ need not be a cycle\footnote{In these considerations we ignore the boundary of the moduli space ${\cal M}$ whose proper treatment requires string field theory \cite{Sen:2014pia}.}; rather, its boundary $\partial \cal{S}$ consists of only ``vertical components", in the sense that
\ie\label{verticalholes}
\partial{\cal S} = \sum_k {\cal Q}_k,
\fe
where each component ${\cal Q}_k$ is a fiber bundle over a chain $\pi({\cal Q}_k) \subset {\cal M}$ of nonzero codimension, whose fiber is of the form $\C\times W \subset (\Sigma\times \overline\Sigma)^{d_o}$, where $\C$ is a 1-cycle in one of the $\Sigma$ or $\overline\Sigma$ factors, and $W$ is a chain in the product of the remaining $\Sigma$ and $\overline\Sigma$ factors.

Under a BRST variation of the string vertex operators, $\widetilde\Omega$ obeys
\ie
\widetilde\Omega[Q_B (\otimes_i {\cal V}_i)] = - d\widetilde\Omega[\otimes_i {\cal V}_i].
\fe
It follows that the BRST variation of the amplitude (\ref{pcoamp}),
\ie
{\cal A}_{g,\epsilon,\widetilde\epsilon}[Q_B (\otimes_i {\cal V}_i)] = - \int_{\cal S} d\widetilde\Omega = - \sum_k\int_{{\cal Q}_k} \widetilde\Omega
\fe
vanishes, since the integral of $\widetilde\Omega$ along each fiber of ${\cal Q}_k$ vanishes by virtue of $\oint_\C d\xi=0$ for 1-cycle $\C\subset \Sigma$ (and similarly for $\C\subset\overline\Sigma$ with $\xi$ replaced by $\widetilde\xi$).

In \cite{Sen:2015hia}, an explicit construction of ${\cal S}$ is given based on a dual triangulation of ${\cal M}$. We can describe a local model of this construction as follows. Let ${\cal D}_1, {\cal D}_2, \cdots, {\cal D}_{2d_e+1}$ be cells of a dual triangulation of ${\cal M}$ that meet at a point. We will denote ${\cal D}_{i_1\cdots i_{p+1}} = {\cal D}_{i_1}\cap \cdots \cap {\cal D}_{i_{p+1}}$ ($i_1<\cdots<i_{p+1}$) the $(2d_e-p)$-dimensional faces along which a subset of the cells meet. The restriction of ${\cal S}$ to the interior of $\pi^{-1}(\cup_i {\cal D}_i)$ is of the form
\ie\label{scontourcomp}
\left.{\cal S}\right|_{{\rm Int}(\pi^{-1}(\cup_i {\cal D}_i))} = \sum_{i_1<\cdots<i_{p+1}} {\cal S}_{i_1\cdots i_{p+1}},
\fe
where ${\cal S}_{i_1\cdots i_{p+1}}$ is a piecewise fiber bundle over ${\cal D}_{i_1\cdots i_{p+1}}$, whose fiber is a $p$-dimensional chain in $(\Sigma\times \overline\Sigma)^{d_o}$ that is the sum of $p$-cubes of the form $\C_1\times\cdots\times \C_p$, where each $\C_i$ is an arc in one of the $\Sigma$ or $\overline\Sigma$ factors. In the $p=0$ case, ${\cal S}_i$ is a section of ${\cal Y}|_{{\cal D}_i}$, which amounts to choosing the locations of $2d_o$ PCOs over the moduli domain ${\cal D}_i$. Moving along ${\cal D}_{i_1\cdots i_{p+1}}$, the fiber ${\cal S}_{i_1\cdots i_{p+1}}$ may jump only by $\C_i\to \C_i + \C'$, where $\C'$ is a 1-cycle in $\Sigma$ or $\overline\Sigma$ (as explained above, such jumps do not affect the integral of $\widetilde\Omega$ along the fiber). Furthermore, the components of (\ref{scontourcomp}) are subject to the matching condition
\ie{}
\left. \sum_{m=1}^{p+1}{\cal S}_{i_1\cdots i_{m-1} i_{m+1}\cdots i_{p+1}}\right|_{{\cal D}_{i_1\cdots i_p}} = - \partial_{ f} {\cal S}_{i_1\cdots i_p},
\fe
where $\partial_f$ stands for taking the boundary of the fiber of ${\cal S}_{i_1\cdots i_p}\to {\cal D}_{i_1\cdots i_p}$.

\section{Integration over the supermoduli space}
\label{sec:integration}

\subsection{Partition of unity}

The integration over a supermanifold $\mathfrak{M}$ of dimension $n_e|n_o$ can be defined via partition of unity, as follows. 
We begin with a sufficiently fine covering of the reduced space ${\cal M}$ with open patches ${\cal U}_\A$, such that each patch ${\cal U}_\A$ can be lifted to a super coordinate patch $\mathfrak{U}_\A$ of ${\mathfrak M}$ via\footnote{Following \cite{Witten:2012bg}, we denote by $\mathbb{R}^{p|*q}$ the complex supersymmetric vector space parameterized by $p$ real bosonic coordinates and $q$ fermionic coordinates, emphasizing that there is no reality condition involving the fermionic coordinates.}
\ie
\varphi_\A: ~{\cal U}_\A\times \mathbb{R}^{0|*n_o} \to \mathfrak{U}_\A.
\fe
Let $\pi_\A: \mathfrak{U}_\A\to {\cal U}_\A$ be the corresponding projection map, namely $\pi_\A\circ\varphi_\A: {\cal U}_\A\times \mathbb{R}^{0|*n_o}\to {\cal U}_\A$ simply forgets the fermionic coordinates. Note that both the lift and projection are highly non-canonical. 
 
Starting with a partition of unity on ${\cal M}$,
\ie
\sum_\A f_\A = 1,~~~~ {\rm Supp} (f_\A) \subset {\cal U}_\A,
\fe
we can construct a partition of unity on $\mathfrak{M}$ via
\ie\label{fapart}
F_\A = {\pi_\A^* f_\A \over \sum_\B \pi_\B^* f_\B},~~~~ \sum_\A F_\A = 1.
\fe
Note that the (fermionic) fibers of the projection $\pi_\A$ and $\pi_\B$ onto two patches ${\cal U}_\A$ and ${\cal U}_\B$ generally do not agree on the overlap ${\cal U}_\A\cup {\cal U}_\B$. Nonetheless, $\left.\sum_\B \pi_\B^* f_\B\right|_{{\mathfrak U}_\A}$ is equal to 1 plus quadratic and higher order terms in the fermionic coordinates and therefore invertible, ensuring that (\ref{fapart}) is well-defined.

The integration of a super form $\Omega$ over $\mathfrak{M}$ is then defined as
\ie
\int_{\mathfrak{M}}\Omega = \sum_\A \int_{{\cal U}_\A\times \mathbb{R}^{0|n_o}} \varphi_\A^*(F_\A\Omega).
\fe
Importantly, the resulting integral does not depend on the choice of partition of unity, and obeys the supermanifold version of Stokes' theorem \cite{Witten:2012bg}. 

\subsection{Integration by lifting and interpolation}
\label{sec:inter}

An alternative way to perform the integration over the supermanifold $\mathfrak{M}$ is to begin with a cell decomposition of the reduced space ${\cal M} = \bigsqcup {\cal D}_\A$, say a dual triangulation, where each domain ${\cal D}_\A$ is contained in a patch ${\cal U}_\A$ that lifts to a super chart $\pi_\A: \mathfrak{U}_\A\to {\cal U}_\A$ as in the previous subsection, perform the integration over each $\pi_\A^{-1}({\cal D}_\A)$, and account for the mismatch between the fibers of $\pi_\A$ and $\pi_\B$ along ${\cal U}_\A\cap {\cal U}_\B$ by an ``interpolating integral". Such a construction amounts to patch-by-patch integration over a top dimensional chain that is homologically equivalent to $\mathfrak{M}$. 

We begin by illustrating the interpolation scheme in a basic example. Consider a supermanifold of real dimension $1|n_o$ covered by two patches $\mathfrak{U}$ and $\widetilde{\mathfrak U}$, with coordinate maps
\ie\label{neighborpatches}
& \varphi: \{ (t, \nu^a) \in \mathbb{R}^{1|*n_o}: t<\delta\} \to^{\!\!\!\!\!\!\!\sim}\, {\mathfrak U},
\\
& \widetilde\varphi:\{ (\widetilde t, \widetilde\nu^a) \in \mathbb{R}^{1|*n_o}: \widetilde t>-\delta\} \to^{\!\!\!\!\!\!\!\sim}\, \widetilde {\mathfrak U} , 
\fe
where $\delta>0$, and the transition map $\widetilde\varphi^{-1}\circ\varphi$ takes the form
\ie\label{transitionmaptf}
\widetilde t = t + f(t, \nu),~~~~ \widetilde \nu^a = g^a(t,\nu),
\fe
for $t\in (-\delta,\delta)$, with $f(t,0)=0$. The reduced space is simply $\mathbb{R}$, covered by ${\cal U} = (-\infty,\delta)$ and $\widetilde{\cal U}=(-\delta,\infty)$. We denote by $\pi: \mathfrak{U}\to {\cal U}$ the projection map such that $\pi\circ \varphi$ simply forgets $\nu^a$, and similarly for $\widetilde\pi: \widetilde{\mathfrak U} \to \widetilde{\cal U}$.

\begin{figure}[h!]
	\centering
	\scalebox{.8}{ 
	\begin{tikzpicture}
		\node[anchor=south west,inner sep=0] (image) at (0,0) {\includegraphics[width=.7\textwidth]{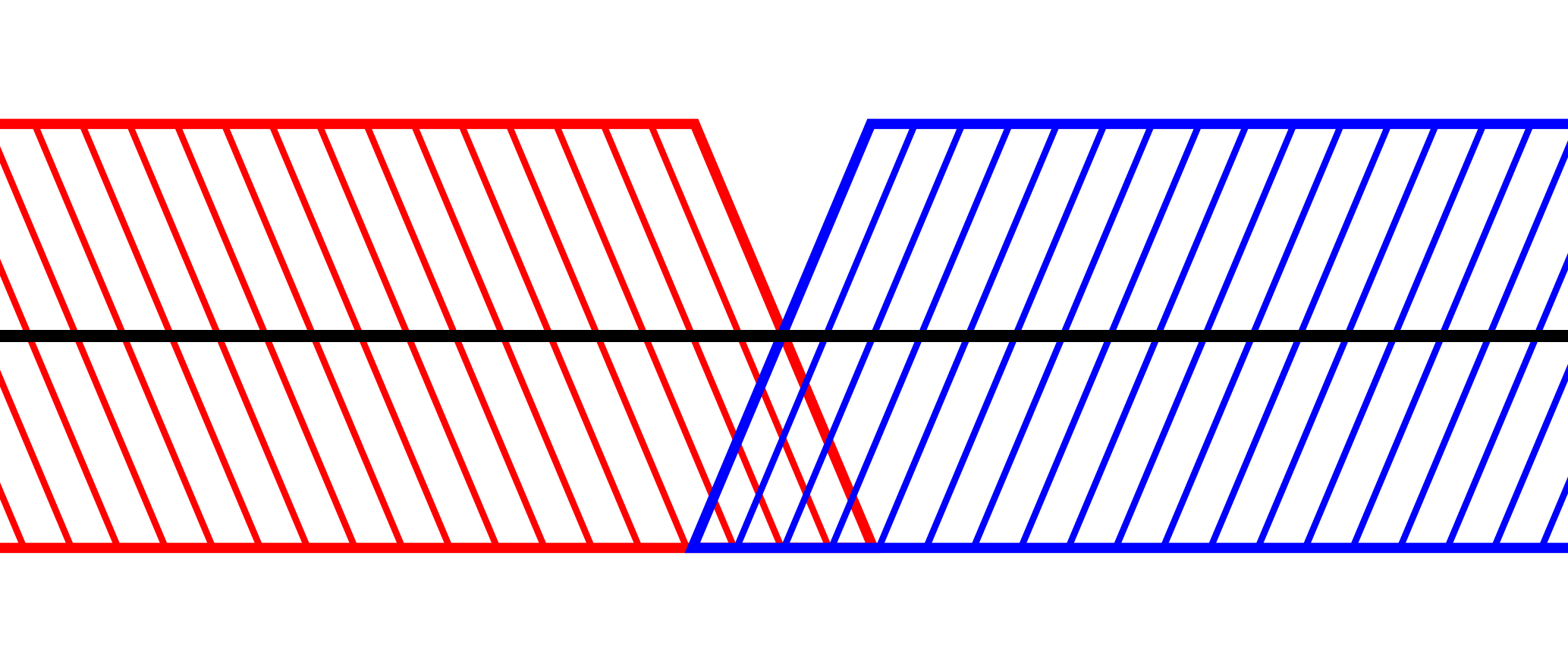}};
		\begin{scope}[shift={(image.south west)}, x={(image.south east)}, y={(image.north west)}]
			\node[anchor=south, color=red] at (0.23, 0.85) {\large $\pi^{-1}(-\infty, 0)$};
			\node[anchor=south, color=blue] at (0.77, 0.85) {\large $\widetilde\pi^{-1}(0, \infty)$};
		\end{scope}
	\end{tikzpicture}}

	\scalebox{.8}{ 	\begin{tikzpicture}
	\node[anchor=south west, inner sep=0] (image2) at (0, 0){\includegraphics[width=0.92\textwidth]{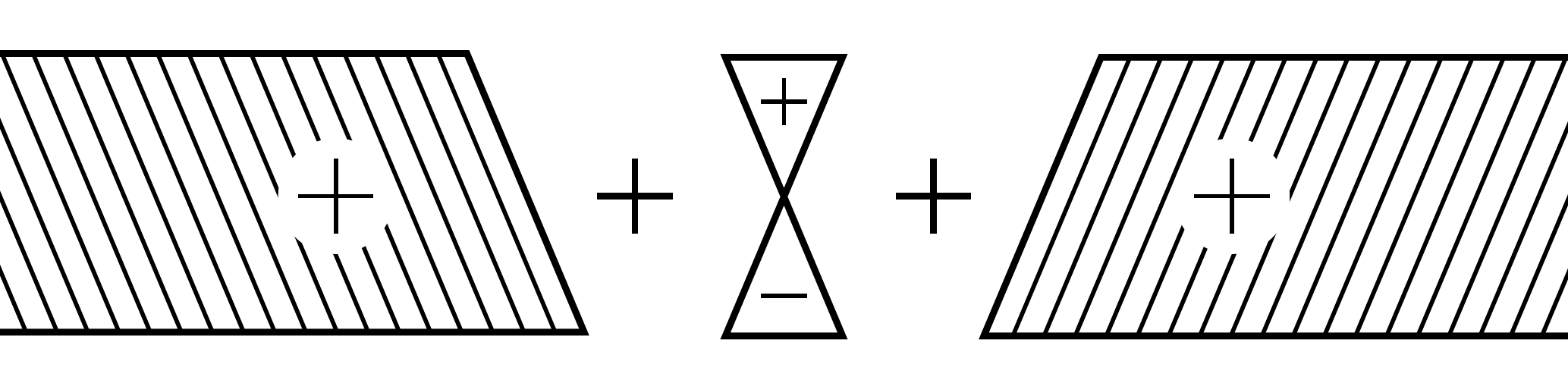}};
	\begin{scope}[shift={(image2.south west)}, x={(image2.south east)}, y={(image2.north west)}]
		\node[anchor=center] at (0.15, 1) {\large $\pi^{-1}(-\infty, 0)$};
		\node[anchor=center] at (0.85, 1) {\large $\widetilde\pi^{-1}(0, \infty)$};
		\node[anchor=center] at (0.5, 1) {\large $\mathfrak{I}_*$};
	\end{scope}
	\end{tikzpicture}
	}
	\caption{A schematic depiction of the interpolation scheme for integrating over $\mathfrak{M}$. The horizontal axis represents the bosonic coordinate $t$ while the vertical direction represents the fermionic coordinates $\nu^a$.
		Top: The mismatched fibers coming from the projections $\pi$ (red, left) and $\widetilde{\pi}$ (blue, right). 
		Bottom: The splitting of the integration contour into integration over the fibers of $\pi$, the fibers of $\widetilde{\pi}$, and the image of the interpolation map $\mathfrak{I}$. 
	}
\end{figure}
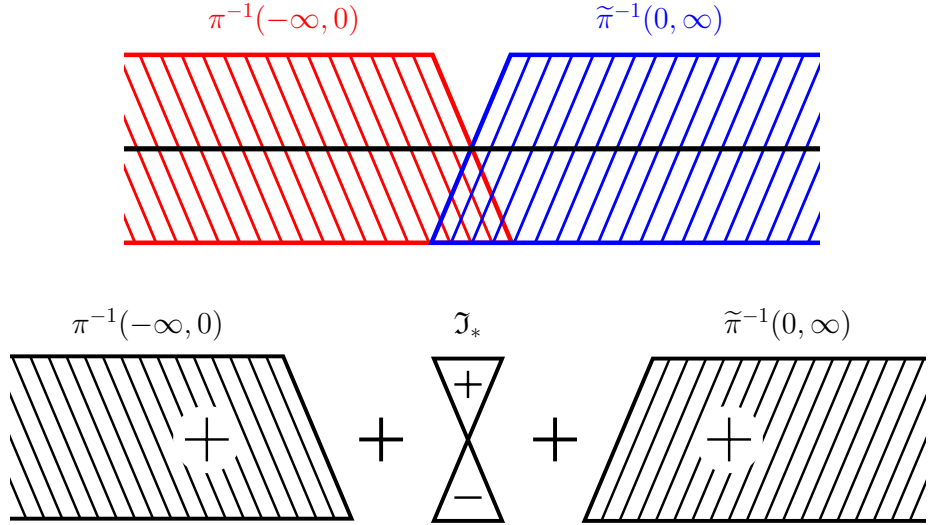

The idea is to break the integration over $\mathfrak{U}\cup \widetilde{\mathfrak U}$ into the integration over $\pi^{-1}(-\infty,0)$ and $\widetilde\pi^{-1}(0,\infty)$, together with an integral over a super patch that interpolates between the fibers $\pi^{-1}(0)$ and $\widetilde\pi^{-1}(0)$. Namely, given a degree $1|n_o$  form $\Omega$, we can write
\ie\label{onedimbreak}
\int_{{\mathfrak U}\cup \widetilde{\mathfrak U}} \Omega = \int_{t<0} \varphi^*\Omega + \int_{\widetilde t>0} \widetilde\varphi^*\Omega + \int_{ [0,1]\times \mathbb{R}^{0|*n_o}} \mathfrak{I}^*\Omega,
\fe
where $\mathfrak{I}$ is an ``interpolation map"
\ie\label{intermap}
& \mathfrak{I} : [0,1]\times \mathbb{R}^{0|*n_o}\to {\mathfrak U}\cap \widetilde{\mathfrak U}
\fe
such that ${\cal J}(\{0\}\times \mathbb{R}^{0|*n_o})$ agrees with $\pi^{-1}(0)$, and that ${\cal J}(\{1\}\times \mathbb{R}^{0|*n_o})$ agrees with $\widetilde\pi^{-1}(0)$. The latter amounts to the matching condition
\ie\label{intermapbdry}
\varphi^{-1}\circ \mathfrak{I}\,(0,\nu) = (0,R(\nu)),~~~~ \widetilde\varphi^{-1}\circ \mathfrak{I}\, (1,\widetilde\nu) = (0, \widetilde R(\widetilde\nu)),
\fe
where $R$ and $\widetilde R$ are super-diffeomorphims from $\mathbb{R}^{0|*n_o}$ to itself. Importantly, we require that $\mathfrak{I}$ has non-singular Berezinian, so that the pullback $\mathfrak{I}^*\Omega$ is a well-defined  form. An explicit construction of the interpolation map $\mathfrak{I}$ and the verification of (\ref{onedimbreak}) is given in Appendix \ref{sec:verifyint}.

To extend this construction to the integration over the supermanifold $\mathfrak{M}$, based on a dual triangulation of its reduced space ${\cal M}$, it suffices to describe a local model of the breakup of the integral. We begin with a local model of dual triangulation in $n_e$ dimensions, where $\mathbb{R}^{n_e}$ is divided into $n_e+1$ domains ${\cal D}_1,\cdots, {\cal D}_{n_e+1}$, defined as
\ie
{\cal D}_i = \{\vec t\in \mathbb{R}^{n_e}: t^k =  \sum_{j\not=i} e_j^k x_j ,~ x_j\geq 0\},
\fe
where $e_j$ are a set of $n_e+1$ vectors in $\mathbb{R}^{n_e}$ that lie on the vertices of a simplex that contains the origin in its interior. We further define the codimension-$p$ faces
\ie
{\cal D}_{i_1\cdots i_{p+1}} = {\cal D}_{i_1}\cap \cdots \cap {\cal D}_{i_p} = \{ \vec t\in \mathbb{R}^{n_e}: t^k = \sum_{j\not=i_1,\cdots,i_{p+1}} e_j^k x_j, x_j\geq 0 \},
\fe
where the subscripts $i_1\cdots i_{p+1}$ are always assumed to be in increasing order. Let
\ie
\iota_j: {\cal D}_{i_1\cdots i_{m-1} j i_m\cdots i_{p}}\to {\cal D}_{i_1\cdots i_{p}}
\fe
be the obvious inclusion map.

Let ${\cal U}_i$ be the open charts containing ${\cal D}_i$,
\ie
{\cal U}_i = \{\vec t\in \mathbb{R}^{n_e}: t^k =  \sum_{j\not=i} e_j^k x_j ,~ x_j\geq -\delta \},
\fe
for some $\delta>0$. Now consider a supermanifold covered by the super charts ${\mathfrak U}_1,\cdots, {\mathfrak U}_{n_e+1}$, each of which admits a coordinate map
\ie
\varphi_i: {\cal U}_i\times \mathbb{R}^{0|*n_o} \to^{\!\!\!\!\!\!\!\sim}\, {\mathfrak U}_i.
\fe
We can then break up the integral of a form $\Omega$ over the supermanifold according to
\ie\label{interform}
\int_{\cup_i {\mathfrak U}_i} \Omega = \sum_{p=0}^{n_e} \sum_{i_1<\cdots< i_{p+1}} \int_{\Delta^{p}\times {\cal D}_{i_1\cdots i_{p+1}} \times \mathbb{R}^{0|*n_o}} \mathfrak{I}_{i_1\cdots i_{p+1}}^*\Omega,
\fe
where $\Delta^p$ is a standard $p$-simplex. The interpolation maps 
\ie{}
& \mathfrak{I}_{i_1\cdots i_{p+1}}: \Delta^{p}\times {\cal D}_{i_1\cdots i_{p+1}} \times \mathbb{R}^{0|*n_o} \to {\mathfrak U}_{i_1}\cap \cdots
\cap {\mathfrak U}_{i_{p+1}}
\fe
are required to have non-singular Berezinian, such that the following matching conditions hold:
\ie\label{intmatchcond}
& \mathfrak{I}_{i_1\cdots i_{p+1}}|_{\partial_m \Delta^{p}} = \mathfrak{I}_{i_1\cdots i_{m-1}i_{m+1}\cdots i_{p+1}} \circ R_{i_1\cdots i_{m-1}i_{m+1}\cdots i_{p+1}}^{(m)} \circ (\sigma_{m,p}\times \iota_{i_m}),~~~p\geq 1,
\\
& \mathfrak{I}_{i} = \varphi_i|_{{\cal D}_i\times \mathbb{R}^{0|*n_o}}.
\fe
Here $\partial_m \Delta^p$ is the $m$-th face ($1\leq m\leq p+1$) of $\Delta^p$, which is identified with the standard $(p-1)$-simplex via the map
\ie
\sigma_{m,p}: \partial_m \Delta^p \to^{\!\!\!\!\!\!\!\sim}\, \Delta^{p-1}.
\fe
$R_{i_1\cdots i_{p}}^{(m)}$ are super-diffeomorphisms on $\Delta^{p-1}\times {\cal D}_{i_1\cdots i_{p}} \times \mathbb{R}^{0|*n_o}$ of the form
\ie
R_{i_1\cdots i_{p}}^{(m)} (\vec s, \vec t, \nu) = (\vec s, \vec t, R_{i_1\cdots i_{p}}^{(m)}(s,t|\nu)), 
\fe
where $R_{i_1\cdots i_{p}}^{(m)}(s,t|\nu)$ have non-singular Berezinian. For our application it will suffice to take $R_{i_1\cdots i_{p}}^{(m)}(s,t|\nu)$ to be a $GL(n_o,\mathbb{C})$ transformation on the $\nu^a$'s, at any given $\vec s\in \Delta^{p-1}$, $\vec t\in {\cal D}_{i_1\cdots i_p}$.

\subsection{Explicit construction of patches on the supermoduli space}
\label{sec:lift}

We will now give an explicit construction of the lifted super charts on the the supermoduli space $\mathfrak{M}_{g,n,\epsilon}$. Let $\{{\cal D}_\A\}$ be a sufficiently fine dual triangulation of the reduced space ${\cal M}_{g,n,\epsilon}$, such that each cell ${\cal D}_\A$ has an open neighborhood ${\cal U}_\A$ parameterizing a family of punctured Riemann surfaces $\Sigma$ equipped with spin structure $\epsilon$, on which we can specify a set of points $\{z_1,\cdots, z_{d_o}\}\subset\Sigma$ that are sufficiently generic in the sense described following (\ref{nvgdef}). Namely, we choose the points $z_a$ to be such that that there is no weight $-{1\over 2}$ meromorphic differential on $\Sigma$ that is compatible with the spin structure $\epsilon$, with only simple poles at $\{z_1,\cdots, z_{d_o}\}$ and simple zeros at the punctures.

The construction of (\ref{superdiscglue}) and (\ref{nvgdef}) then defines a family of SRS $\mathfrak{C}_\nu$ whose reduced space is $\Sigma$ with spin structure $\epsilon$, that depend on the fermionic parameters $\nu^a$, $a=1,\cdots,d_o$. Said equivalently, this construction give the coordinate map of a super chart ${\mathfrak U}_\A \subset \mathfrak{M}_{g,n,\epsilon}$,
\ie\label{phiamap}
\varphi_\A: {\cal U}_\A \times \mathbb{R}^{0|*d_o}\to {\mathfrak U}_\A ,
\fe
that takes $(\Sigma,\epsilon)\in{\cal U}_\A$ and $\nu\in\mathbb{R}^{0|*d_o}$ to $\mathfrak{C}_\nu$. Importantly, the genericness of $\{z_a\}$ in the sense defined above ensures that $\varphi_\A$ is non-degenerate. As before we will write $\pi_\A: \mathfrak{U}_\A \to {\cal U}_\A$ as the projection map, such that $\pi_\A\circ \varphi_\A^{-1}$ simply forgets the odd parameters. 

On the overlap between two super charts ${\mathfrak U}_\A$ and ${\mathfrak U}_\B$, the coordinate maps $\varphi_\A$ and $\varphi_\B$ are constructed by choosing two different sets of points $\{z_a\}$ and $\{z_a' \}$ on $\Sigma$. To exhibit the transition map $\varphi_\B^{-1}\circ \varphi_\A$, it suffices to consider the case where $z_a'=z_a$ for $a=2,\cdots, d_o$, and only $z_1'$ differs from $z_1$, as follows.

The SRS that corresponds to $\varphi_\A(\Sigma,\epsilon,\nu)$ (omitting the anti-holomorphic data) is obtained from the split SRS $\mathfrak{C}_0$ by the insertion of
\ie\label{nwgdef}
\prod_{a=1}^{d_o} \left[1+ \nu^a G(z_a) \right] .
\fe
There is a weight $-{1\over 2}$ meromorphic differential $\zeta = f(z) (dz)^{-{1\over 2}}$ on $\Sigma$ (see (\ref{meroweighthalf})), with $d_o+1$ poles at $z_1, z_2, \cdots, z_{d_o}$ and $z_1'$, as well as simple zeros at each puncture of $\Sigma$, normalized such that the residue of $f(z)$ at $z=z_1$ is equal to 1. This allows us to express the insertion of $G(z_1)$ as a contour integral,
\ie\label{gcont}
G(z_1) &= \oint_{C_{z_1}} {dz\over 2\pi i}  f(z) G(z),
\fe
where $C_{z_1}$ is a counterclockwise contour that encloses $z_1$. We will denote by $f^{(n)}(w)$ the coefficient of $(z-w)^n$ in the Laurent series of $f(z)$ around $z=w$ (which can be nonzero for $n\geq -1$). (\ref{nwgdef}) can be written equivalently as
\ie\label{gprodinter}
& \left[ 1 + \nu^1 \oint_{C_{z_1}} {dz\over 2\pi i}  f(z) G(z) \right] \prod_{a=2}^{d_o} \left[1+ \nu^a G(z_a) \right] 
\\
&= \left[ 1 - \nu^1 f^{(-1)}(z_1') G(z_1') - \nu^1 \sum_{b=2}^{d_o} \oint_{C_{z_b}} {dz\over 2\pi i}  f(z) G(z) \right] \prod_{a=2}^{d_o} \left[1+ \nu^a G(z_a) \right] 
\\
&= \left[ 1 - \nu^1 f^{(-1)}(z_1') G(z_1')  \right] \prod_{a=2}^{d_o} \left[1+ \nu^a G(z_a) \right] 
\\
&~~~~ - \nu^1\sum_{b=2}^{d_o} \left[ f^{(-1)}(z_b) G(z_b) - 2 \nu^b f^{(0)}(z_b) T(z_b) \right] \prod_{a=2, a\not=b}^{d_o} \left[1+ \nu^a G(z_a) \right] ,
\fe
where in deriving the last equality we have used (\ref{ggope}) and the absence of order $z^0$ term in the $G(z) G(0)$ OPE. We can then put (\ref{gprodinter}) back in the product form,
\ie
\left[ 1+ \nu'^1 G(z_1')  \right] \prod_{a=2}^{d_o} \left[1+ \nu'^a G(z_a) \right] \left[1 + \sum_{b=2}^{d_o} 2 \nu^1 \nu^b  f^{(0)}(z_b) T(z_b)  \right],
\fe 
where
\ie{}
& \nu'^1 = - \nu^1 f^{(-1)}(z_1'),
\\
& \nu'^a = \nu^a - \nu^1 f^{(-1)}(z_a) ,~~~a=2,\cdots,d_o.
\fe
From this we read off the transition map $\varphi_\B^{-1}\circ \varphi_\A$ on $({\cal U}_\A\cap {\cal U}_\B)\times \mathbb{R}^{0|*d_o}$,
\ie
\varphi_\B^{-1}\circ \varphi_\A: (\tau^m,\nu^a) \mapsto ( \tau'^m = \tau^m + \delta\tau^m, \nu'^a),
\fe
where the bosonic moduli shift $\delta\tau^m$ is generated by the insertion of 
\ie
\sum_{b=2}^{d_o} 2 \nu^1 \nu^b  f^{(0)}(z_b) T(z_b). 
\fe
By moving points of $\{z_a\}$ one at a time, we can obtain the general transition maps between arbitrary pairs of super charts $\mathfrak{U}_\A$.

\subsection{The supermoduli integration contour $\mathfrak{S}$}
\label{sec:supercontour}

As explained in section \ref{sec:superamp}, we will be integrating not over the supermoduli space itself, but rather a complex codimension $d_e|0$ contour $\mathfrak{S} \subset \mathfrak{M}_{g,n,\epsilon} \times \overline{{\mathfrak M}_{g,n,\widetilde\epsilon}}$, whose reduced space is the diagonal subspace ${\cal M}\subset {\cal M}_{g,n,\epsilon} \times \overline{{\cal M}_{g,n,\widetilde\epsilon}}$. Different choices of $\mathfrak{S}$ are equivalent in homology, and result in the same integral of the form $\Omega$ which is closed for on-shell amplitudes.

Given a sufficiently fine dual triangulation $\{{\cal D}_\A\}$ of ${\cal M}$ and open patches ${\cal U}_\A\supset {\cal D}_\A$, we can construct the lifting map $\varphi_\A$ defined as in (\ref{phiamap}), based on the choice of a set of points $\{z_a^{(\A)} \}\subset \Sigma$, with $\varphi_\A({\cal U}_\A\times \mathbb{R}^{0|*d_o}) = \mathfrak{U}_\A\subset \mathfrak{M}_{g,n,\epsilon}$, and $\pi_\A(\mathfrak{U}_\A) = {\cal U}_\A$. Similarly, we can define the anti-holomorphic lifting map $\overline\varphi_\A$ that depends on $\{\bar z_a^{(\A)} \}\subset \overline\Sigma$, and the corresponding super chart $\overline{\mathfrak U}_\A\subset \overline{{\mathfrak M}_{g,n,\widetilde\epsilon}}$. Note that $\bar z_a^{(\A)}$ need not be complex conjugate to $z_a^{(\A)}$. Applying both the holomorphic and anti-holomorphic lifting map to a cell ${\cal D}_\A$ produces a patch
\ie\label{phipatch}
(\varphi_\A\times \overline\varphi_\A ) ({\cal D}_\A\times \mathbb{R}^{0|*d_o} \times \mathbb{R}^{0|*d_o}) = (\pi_\A^{-1}\times \overline \pi_\A^{-1}) ({\cal D}_\A)
\fe
of the contour $\mathfrak{S}$, which we will refer to as a ``horizontal" patch.

Next, we would like to construct interpolating ``vertical" patches that fill in the gaps between horizontal patches of the form (\ref{phipatch}) and close the chain $\mathfrak{S}$. This is analogous to (\ref{interform}) but applied to integration over a submanifold. Suppose the cells ${\cal D}_{\A_1},\cdots,{\cal D}_{\A_{p+1}}$ of the dual triangulation of ${\cal M}$ meet along the codimension-$p$ face ${\cal D}_{\A_1\cdots \A_{p+1}}$. We aim to construct a set of interpolation maps 
\ie\label{intcontoumaps}
\mathfrak{I}_{\A_1\cdots \A_{p+1}}: \Delta^p\times {\cal D}_{\A_1\cdots \A_{p+1}}\times \mathbb{R}^{0|*d_o} \times \mathbb{R}^{0|*d_o} \to {\mathfrak U}_{\A_1\cdots \A_{p+1}} \times \overline{\mathfrak U}_{\A_1\cdots \A_{p+1}},
\fe
where ${\mathfrak U}_{\A_1\cdots \A_{p+1}}\equiv {\mathfrak U}_{\A_1} \cap\cdots \cap{\mathfrak U}_{\A_{p+1}}$, that obey matching conditions analogous to (\ref{intmatchcond}), of the form
\ie\label{supcontr}
&  \mathfrak{I}_{\A_1\cdots \A_{p+1}}|_{\partial_m \Delta^{p}} = \mathfrak{I}_{\A_1\cdots \A_{m-1}\A_{m+1}\cdots \A_{p+1}} \circ R_{\A_1\cdots \A_{m-1}\A_{m+1}\cdots \A_{p+1}}^{(m)} \circ (\sigma_{m,p}\times \iota_{i_m}),~~~p\geq 1,
\\
& {\mathfrak I}_\A =\left. \varphi_\A\times \overline\varphi_\A \right|_{{\cal D}_\A\times \mathbb{R}^{0|*d_o} \times \mathbb{R}^{0|*d_o}}.
\fe
The contour $\mathfrak{S}$ is then built as
\ie
\mathfrak{S} = \sum_{p=0}^{2d_e} \sum_{\{\A_1,\cdots,\A_{p+1}\}} \mathfrak{I}_{\A_1\cdots \A_{p+1}}( \Delta^p\times {\cal D}_{\A_1\cdots \A_{p+1}}\times \mathbb{R}^{0|*d_o} \times \mathbb{R}^{0|*d_o}).
\fe
In the next section, we will describe a specific construction of (\ref{intcontoumaps}) that leads to precisely the vertical integration prescription.

\section{From the supermoduli contour to the PCO contour}
\label{sec:mtos}

\subsection{PCO from cutting a super disc}
\label{sec:pcoemerge}

We begin by considering the effect of integrating out a single fermionic modulus on a patch of the supermoduli space that parameterizes the gluing map of a super disc according to (a slight generalization of) (\ref{superdiscglue}). Namely, consider a super disc $D$ with coordinates $(w,\eta)$, and an overlapping super annulus $U'$ with coordinates $(z,\theta)$, with the transition map 
\ie\label{transidisc}
& w = z - \frac{\theta \nu}{z - z_0(t)} ,
\\
& \eta = \theta - \frac{\nu}{z - z_0(t)}.
\fe
Here $\nu$ is an odd parameter, and the $z_0(t)$ can have arbitrary dependence on the bosonic moduli $t^k=(\tau^m, \bar\tau^m)$.

According to (\ref{nvgdef}) and (\ref{omegaintg}), the transition map (\ref{transidisc}) gives rise to the following $\nu$-dependent factor in the supermoduli integrand,
\ie\label{bbfact}
\left[ 1 + \nu \oint {dz\over 2\pi i} {G(z)\over z-z_0(t)} \right]  \exp\left[ {\cal B}_{z_0(t)} \right] \delta(d\nu) \delta({\cal B}_\nu) ,
\fe
where the $z$-integral contour encircles $z_0(t)$. ${\cal B}_{z_0(t)}$ and ${\cal B}_\nu$ are given by
\ie
{\cal B}_{z_0(t)} &= \oint {dw\over 2\pi i} d\eta\, (\B + \eta b)\left. \left[ {\partial w\over \partial t^k} - {\partial\eta\over \partial t^k}\eta \right]\right|_{z,\theta} dt^k 
\\
&= - 2 \nu \partial\beta(z_0(t)) dz_0(t) 
\fe
and
\ie
{\cal B}_\nu &= \oint {dw\over 2\pi i} d\eta\, (\B + \eta b)\left. \left[ {\partial w\over \partial\nu} - {\partial\eta\over \partial\nu}\eta \right]\right|_{z,\theta}
\\
&= \oint {dw\over 2\pi i} d\eta\, (\B + \eta b) \left[ {2\eta\over w-z_0(t)} + {\nu\over (w-z_0(t))^2} \right]
\\
&= 2\B(z_0(t)) - \nu \partial b(z_0(t))
\fe
respectively. Thus, (\ref{bbfact}) can be written as
\ie{}
&\left[ 1 + \nu \oint {dz\over 2\pi i} {G(z)\over z-z_0(t)} \right] \Big[ 1 - 2\nu  \partial\beta(z_0(t)) dz_0(t)  \Big] \cdot \delta(d\nu)\left[ {1\over 2} \delta(\B(z_0(t))) - {1\over 4} \nu \partial b(z_0(t)) \,\delta'(\B(z_0(t))) \right] 
\fe
Integrating this over $\nu$ gives
\ie{}
& {1\over 2}\oint \frac{\dif z}{2 \pi i} \frac{G(z)}{z - z_0(t)} \delta(\beta(z_0(t))) - \frac{1}{4}\partial b(z_0(t)) \,\delta'(\beta(z_0(t))) - dz_0(t)  \partial\Theta(\B(z_0(t))) 
\\
&= {\cal X}(z_0(t)) + d\xi(z_0(t)),
\fe
where $\xi=-\Theta(\B)$, and 
\ie
{\cal X}(z) = {1\over 2}\oint {dz'\over 2\pi i} {G(z')\over z'-z} \delta(\B(z)) - {1\over 4} \partial b(z) \delta'(\B(z)) 
\fe
is precisely the PCO.

Applying this to all of the odd moduli parameters in the integration over the lifted patch $(\pi_\A^{-1}\times \overline\pi_\A^{-1})({\cal D}_\A)$ of section \ref{sec:lift}, where the points $\{z_a^{(\A)}\}\subset \Sigma$ and $\{\bar z_a^{(\A)}\}\subset \overline\Sigma$ are chosen to be independent of the local bosonic moduli parameter $t^k$, we obtain
\ie\label{tmpdd}
\int_{(\pi_\A^{-1}\times \overline\pi_\A^{-1})({\cal D}_\A)} \Omega = \int_{{\cal D}_\A} \left\langle e^{\cal B}\prod_{i=1}^n {\cal V}_i  \prod_{a=1}^{d_o} \mathcal{X}(z_a^{(\A)}) \widetilde{\mathcal{X}}(\bar z_a^{(\A)})  \right\rangle.
\fe
This amounts to a ``horizontal patch" of the integration over the PCO contour (\ref{pcoamp}). We could of course slightly generalize (\ref{tmpdd}) to allow $z_a$ and $\bar z_a$ to depend on $t$, and recover an expression that involves the integrand (\ref{tildeomg}) including the $d\xi$, $d\widetilde\xi$ terms, but this is inessential for our consideration. In fact, simply allowing $z_a, \bar z_a$ to depend on $t$ would not enable us to close the chain $\mathfrak{S}$, as we must ensure that the former evade configurations that lead to spurious singularities. Instead, we will proceed by constructing the interpolating patches in the supermoduli space.

\subsection{The emergence of vertical integration}
\label{sec:vertemerge}

We now construct the interpolation map $\mathfrak{I}_{\A\B}$ between a pair of neighboring cells ${\cal D}_{\A},{\cal D}_{\B}\subset{\cal M}$ that meet along the codimension 1 face ${\cal D}_{\A\B}$, as the $p=1$ case of (\ref{intcontoumaps}). The lifting maps $\varphi_\A, \overline{\varphi}_\A$ are defined as in section \ref{sec:supercontour}, involving the choice of points $\{z_a^{(\A)}\}\subset\Sigma$, $\{\overline z_a^{(\A)}\}\subset\overline\Sigma$, and similarly for $\varphi_\B, \overline{\varphi}_\B$ with the points $\{z_a^{(\B)}\}$, $\{\bar z_a^{(\B)}\}$. $\mathfrak{I}_{\A\B}$ will be constructed as a composition (in the sense of homotopy) of a sequence of $2d_o$ interpolation maps 
\ie\label{iibar}
& {\cal I}_{\bf a}: [0,1]\times {\cal D}_{\A\B} \times\mathbb{R}^{0|*d_o}\to {\mathfrak U}_{\A\B},
\\
& \overline{\cal I}_{\bf a}: [0,1]\times {\cal D}_{\A\B} \times\mathbb{R}^{0|*d_o}\to \overline{\mathfrak U}_{\A\B},
\fe
according to
\ie
& \mathfrak{I}_{\A\B}(s,t,\nu,\overline\nu) = ({\cal I}_{\bf a}(2d_o s-{\bf a}+1, t, \nu) , \overline\varphi_\A(t,\overline\nu)),~~~~{{\bf a}-1\over 2d_o} \leq s\leq {{\bf a}\over 2d_o},
\\
& \mathfrak{I}_{\A\B}(s,t,\nu,\overline\nu) = (\varphi_\B(t,\nu), \overline{\cal I}_{\bf a}(2d_o s-d_o-{\bf a}+1, t, \overline\nu) ),~~~~{{\bf a}-1\over 2d_o}+{1\over 2} \leq s\leq {{\bf a}\over 2d_o}+{1\over 2},
\fe
for ${\bf a}=1,\cdots,d_o$.

${\cal I}_{\bf a}(s,t,\nu)$ is defined as the SRS obtained from the split SRS $\mathfrak{C}_0$ over $(\Sigma_t,\epsilon)$ by 
cutting $d_o+1$ super discs centered at
\ie\label{zloci}
z_1^{(\B)}, z_2^{(\B)}, \cdots, z_{\bf a}^{(\B)}, z_{\bf a}^{(\A)}, z_{{\bf a}+1}^{(\A)},\cdots, z_{d_o}^{(\A)}.
\fe
The gluing maps at the boundary of the first ${\bf a}-1$ discs, centered at $z_1^{(\B)},\cdots, z_{{\bf a}-1}^{(\B)}$, are the same as those that define the lifting map $\varphi_\B$, namely (\ref{superdiscglue}) with $z_a$ replaced with $z_a^{(\B)}$, that depend on the Grassmann-odd parameters $\nu^1,\cdots,\nu^{{\bf a}-1}$. The gluing maps at the boundary of the last $d_o-{\bf a}$ discs, centered at $z_{{\bf a}+1}^{(\A)},\cdots, z_{d_o}^{(\A)}$, are the same as those that define the lifting map $\varphi_\A$, that depend on $\nu^{{\bf a}+1},\cdots,\nu^{d_o}$. On the boundary of the discs centered at $z_{\bf a}^{(\B)}$ and $z_{\bf a}^{(\A)}$, however, we use a gluing map that interpolates between those that define $\varphi_\A$ and $\varphi_\B$ as follows.

Let $(w,\eta)$ be the coordinates on the super disc centered at $z_{\bf a}^{(\A)}$, and $(\widetilde w,\widetilde\eta)$ be the coordinates on the super disc centered at $z_{\bf a}^{(\B)}$. The gluing maps that depend on $\nu^a$ are
\ie\label{wetag}
w=z-{\theta (1-s) R(s) \nu^{\bf a}\over z- z_{\bf a}^{(\A)}},~~~~\eta = \theta - {(1-s) R(s)\nu^{\bf a}\over z- z_{\bf a}^{(\A)}},
\fe
and
\ie\label{wetatil}
\widetilde w=z-{\theta s \widetilde R(s)\nu^{\bf a}\over z- z_{\bf a}^{(\B)}},~~~~\widetilde\eta = \theta - {s \widetilde R(s)\nu^{\bf a}\over z- z_{\bf a}^{(\B)}},
\fe
where $R(s)$ and $\widetilde R(s)$ are nowhere vanishing complex valued functions on $[0,1]$. Note that our matching conditions do not require $R(0)$ and $\widetilde R(1)$ to be identity.

To first order in $\nu$, the deformation that defines the SRS ${\cal I}_{\bf a}(s,t,\nu)$ amounts to the insertion of
\ie\label{sumgs}
\sum_{b=1}^{{\bf a}-1} \nu^b G(z_b^{(\B)}) + \sum_{b'={\bf a}+1}^{d_o} \nu^{b'} G(z_{b'}^{(\A)}) + \nu^{\bf a}  \left[ (1-s) R(s) G(z_{\bf a}^{(\A)}) + s \widetilde R(s) G(z_{\bf a}^{(\B)})\right]
\fe
on the worldsheet. For generic $\{z_a^{(\A)}\}$ and $\{z_a^{(\B)}\}$, up to constant rescaling there is a unique weight $-{1\over 2}$ meromorphic differential $\zeta = f(z) (dz)^{-{1\over 2}}$ with $d_o+1$ simple poles at (\ref{zloci}). By consideration of contour deformation similarly to that at the end of section \ref{sec:srsscft}, the variation of (\ref{sumgs}) under $\nu\to \nu+\delta\nu$ is degenerate if 
\ie\label{srresid}
{s \widetilde R(s) \over (1-s) R(s)} = {f^{(-1)}(z_{\bf a}^{(\B)}) \over f^{(-1)} (z_{\bf a}^{(\A)})}
\fe
where $f^{(-1)}(z_0)$ stands for the residue of $f(z)$ at $z=z_0$, for some $s\in[0,1]$. With generic choices of complex $R(s)$ and $\widetilde R(s)$, (\ref{srresid}) can certainly be avoided, thereby ensuring that ${\cal I}_a$ has non-singular Berenzinian, as needed.

The anti-holomorphic counter part to ${\cal I}_{\bf a}$, which we denoted by $\overline{\cal I}_{\bf a}$ in (\ref{iibar}), is defined similarly via gluing maps on the complex conjugate SRS. Note that the points $\bar z_a^{(\A)}, \bar z_a^{(\B)}\in \overline\Sigma$ involved in this construction need not be the complex conjugates of $z_a^{(\A)}, z_a^{(\B)}\in\Sigma$.

The integration over the interpolation patch defined by $\mathfrak{I}_{\A\B}$ can be written as
\ie\label{iabint}
\int_{[0,1]\times {\cal D}_{\A\B}\times \mathbb{R}^{0|*d_o}\times \mathbb{R}^{0|*d_o}} \mathfrak{I}_{\A\B}^*\Omega = \int_{[0,1]\times {\cal D}_{\A\B}\times \mathbb{R}^{0|*d_o}\times \mathbb{R}^{0|*d_o}} \left[ \sum_{{\bf a}=1}^{d_o} ({\cal I}_{\bf a} \times \overline\varphi_\A)^*\Omega + \sum_{{\bf a}=1}^{d_o} (\varphi_\B\times\overline{\cal I}_{\bf a})^*\Omega \right]
\fe
The integrand $({\cal I}_{\bf a} \times \overline\varphi_\A)^*\Omega$, for instance, contains the following $s$ and $\nu^{\bf a}$ dependent factor
\ie\label{bbsfact}
{\cal B}_s \delta(d\nu^{\bf a}) \delta({\cal B}_{\nu^{\bf a}}),
\fe
where
\ie{}
& {\cal B}_s = - 2 \nu^{\bf a} \left[ \B(z_{\bf a}^{(\A)}) {d\over ds}((1-s)R(s)) + \B(z_{\bf a}^{(\B)}) {d\over ds}(s \widetilde R(s)) \right] ds
\\
& {\cal B}_{\nu^{\bf a}} = 2\left[ (1-s) R(s)\B(z_{\bf a}^{(\A)}) + s\widetilde R(s) \B(z_{\bf a}^{(\B)})\right] - \nu^{\bf a} \left[ ((1-s) R(s))^2 \partial b(z_a^{(\A)}) + (s\widetilde R(s))^2 \partial b( z_{\bf a}^{(\B)}) \right].
\fe
Note that we don't need to include insertion of $G$ associated with the $\nu^{\bf a}$-deformation in (\ref{bbsfact}) because ${\cal B}_s$ already saturates the $\nu^{\bf a}$-integral.

Integrating (\ref{bbsfact}) over $s$ and $\nu^{\bf a}$ results in
\ie\label{thetadiff}
&- \int_0^1 ds \left[ \B(z_{\bf a}^{(\A)}) {d\over ds}((1-s)R(s)) + \B(z_{\bf a}^{(\B)}) {d\over ds}(s \widetilde R(s)) \right]\delta\left( (1-s) R(s)\B(z_{\bf a}^{(\A)}) + s\widetilde R(s) \B(z_{\bf a}^{(\B)}) \right)
\\
&= -\int_0^1 ds\, \partial_s \Theta\left( (1-s) R(s)\B(z_{\bf a}^{(\A)}) + s\widetilde R(s) \B(z_{\bf a}^{(\B)}) \right)
\\
&= -\Theta\big(\widetilde R(1) \B(z_{\bf a}^{(\B)})\big) + \Theta\big(R(0) \B(z_{\bf a}^{(\A)})\big).
\fe
Recall that the holomorphic distribution $\Theta(\B)$ is homogenous in its argument, and thus the RHS of (\ref{thetadiff}) is the same as
\ie
\xi(z_{\bf a}^{(\B)}) - \xi(z_{\bf a}^{(\A)}).
\fe
This matches precisely the vertical integration prescription associated with moving a holomorphic PCO from $z_{\bf a}^{(\A)}$ to $z_{\bf a}^{(\B)}$. 

(\ref{iabint}) is therefore equivalent to the sum over $2d_o$ vertical integrals, each of which involves moving a single holomorphic or anti-holomorphic PCO. The result can be written as
\ie{}
&\int_{[0,1]\times {\cal D}_{\A\B}\times \mathbb{R}^{0|*d_o}\times \mathbb{R}^{0|*d_o}} \mathfrak{I}_{\A\B}^*\Omega 
\\
&=\int_{{\cal D}_{\A\B}}  \Bigg\langle e^{\cal B}  \prod_{i=1}^n {\cal V}_i  \Bigg[  \sum_{{\bf a}=1}^{d_o} \prod_{b=1}^{{\bf a}-1} {\cal X}(z_b^{(\B)}) \big[ \xi(z_{\bf a}^{(\B)}) - \xi(z_{\bf a}^{(\A)}) \big]   \prod_{b'={\bf a}+1}^{d_o} {\cal X}(z_{b'}^{(\A)}) \prod_{c=1}^{d_o} \widetilde{\cal X}(\bar z_c^{(\A)})
\\
&~~~~~~~~~~~~~~~ + \sum_{{\bf a}=1}^{d_o}  \prod_{b=1}^{d_o}{\cal X}(z_b^{(\B)}) \prod_{c=1}^{{\bf a}-1} \widetilde{\cal X}(\bar z_c^{(\B)}) \big[ \widetilde\xi(\bar z_{\bf a}^{(\B)}) - \widetilde\xi(\bar z_{\bf a}^{(\A)}) \big]   \prod_{c'={\bf a}+1}^{d_o} \widetilde{\cal X}(\bar z_{c'}^{(\A)})   \bigg] \Bigg\rangle.
\fe

\subsection{Recovering the full PCO contour ${\cal S}$}
\label{sec:pcohighercodim}


To complete the supermoduli integration contour $\mathfrak{S}$ outlined in section \ref{sec:supercontour}, we now extend the construction of the interpolation map of the previous subsection to $\mathfrak{I}_{\A_1 \cdots \A_{p+1}}$ supported at the codimension $p$ interface between ${\cal D}_{\A_1},\cdots,{\cal D}_{\A_{p+1}}\subset {\cal M}$. Each cell ${\cal D}_{\A_m}$ is contained in its open neighborhood ${\cal U}_{\A_m}$, the latter being lifted to a supermoduli patch $\mathfrak{U}_{\A_m}$ via the map $\varphi_{\A_m}$ specified through PCO locations $\{z_a^{(\A_1)}\},\cdots,\{z_a^{(\A_{p+1})}\}$ in the sense of section \ref{sec:pcoemerge}, along with their anti-holomorphic counter parts.

We begin with an auxiliary space 
\ie
V = \Delta_1\times\cdots\times\Delta_{d_o},
\fe
where each $\Delta_a$ is a $p$-simplex, whose vertices are denoted $\{v_{a,1},\cdots, v_{a,p+1}\}$. We will refer to a tuple of vertices $(v_{1,x_1},\cdots,v_{d_o,x_{d_o}})\in V$, where $1\leq x_a\leq p+1$, as a vertex of $V$. Such vertices are in 1-1 correspondence with ``shuffled" PCO locations $\{z_a^{(\A_{x_a})}\}_{1\leq a\leq d_o}$, where the $a$-th PCO position is taken from that of the lifting map $\varphi_{\A_{x_a}}$. A sequence of PCO moves from $\{z_a^{(\A_{x_a})}\}$ to $\{z_a^{(\A_{y_a})}\}$ can be associated with a path along the edges of the embedded $d_o$-hypercube $\prod_{a=1}^{d_o}\overline{v_{a,x_a} v_{a,y_a}} \,\subset V$. For instance,  the interpolation map $\mathfrak{I}_{\A_x \A_y}$ of section \ref{sec:vertemerge} is associated with the path
\ie
\Gamma_{xy}: (v_{1,x}, v_{2,x},\cdots,v_{d_o,x})\to (v_{1,y}, v_{2,x},\cdots,v_{d_o,x})\to \cdots \to (v_{1,y}, v_{2,y},\cdots,v_{d_o,y}),
\fe
which shall be viewed as a piecewise linear map from the standard 1-simplex to $V$.

Given three such paths, say $\Gamma_{\A_1\A_2}$, $\Gamma_{\A_2\A_3}$, $\Gamma_{\A_1\A_3}$ associated with $\mathfrak{I}_{\A_1\A_2}$, $\mathfrak{I}_{\A_2\A_3}$, $\mathfrak{I}_{\A_1\A_3}$, we can form a closed path as the piecewise linear map
\ie\label{gammabdry}
\partial\Gamma_{\A_1\A_2\A_3}: \partial\Delta^2 \to V
\fe
composed by joining the three paths, each of which is defined on a boundary segment $\partial_m \Delta^2$, $m=1,2,3$. Evidently, $\partial\Gamma_{\A_1\A_2\A_3}$ is contractible and can be extended to 
\ie\label{gammathree}
\Gamma_{\A_1\A_2\A_3}: \Delta^2 \to V.
\fe
It is convenient to realize $\Gamma_{\A_1\A_2\A_3}$ as a homotopy between the path $\Gamma_{\A_1\A_2}\triangleright \Gamma_{\A_2\A_3}$ (here $\triangleright$ stands for ``joining") and $\Gamma_{\A_1\A_3}$, that is composed of a sequence of elementary homotopies of two types.

The first type of elementary homotopy, which we refer to as a ``merger", is a homotopy between two paths $\Gamma$ and $\Gamma'$ that differ only at the respective segments
\ie\label{mergeraa}
&  v_{{\bf a},1} \to v_{{\bf a},2} \to v_{{\bf a},3}  \subset \Gamma,
\\
&  v_{{\bf a},1} \to v_{{\bf a},3} \subset \Gamma',
\fe
where we have exhibited only the movement between vertices in $\Delta_{\bf a}$, while the coordinates in each of the  other simplex factors of $V$, namely $\Delta_1,\cdots,\Delta_{{\bf a}-1},\Delta_{{\bf a}+1},\cdots,\Delta_{d_o}$, are fixed to be at some vertex. The merger is defined as the homotopy associated with the obvious map from $\Delta^2$ to the 2-simplex spanned by $\{v_{{\bf a},1}, v_{{\bf a},2},v_{{\bf a},3}\}$ in $\Delta_{\bf a}$.

The second type of elementary homotopy, which we refer to as a ``square move", is a homotopy between $\Gamma$ and $\Gamma'$ that differ at the following segments,
\ie\label{squareab}
& (v_{{\bf a},2}, v_{{\bf b},1}) \to (v_{{\bf a},2}, v_{{\bf b},2}) \to (v_{{\bf a},3}, v_{{\bf b},2})  \subset\Gamma,
\\
& (v_{{\bf a},2}, v_{{\bf b},1}) \to (v_{{\bf a},3}, v_{{\bf b},1}) \to (v_{{\bf a},3}, v_{{\bf b},2}) \subset\Gamma',
\fe
where we have exhibited only the movement of a pair of vertices on $\Delta_{\bf a}$ and $\Delta_{\bf b}$, where $1\leq a<b\leq d_o$, while the coordinates on the other $\Delta_c$ factors for $c\not={\bf a},{\bf b}$ are at some fixed vertices along these segments of the paths. The square move is defined as the homotopy associated with the obvious map from $[0,1]^2$ to the square $\overline{v_{{\bf a},2} v_{{\bf a},3}}\times \overline{v_{{\bf b},1}v_{{\bf b},2}}$ in $\Delta_{\bf a}\times \Delta_{\bf b}$.

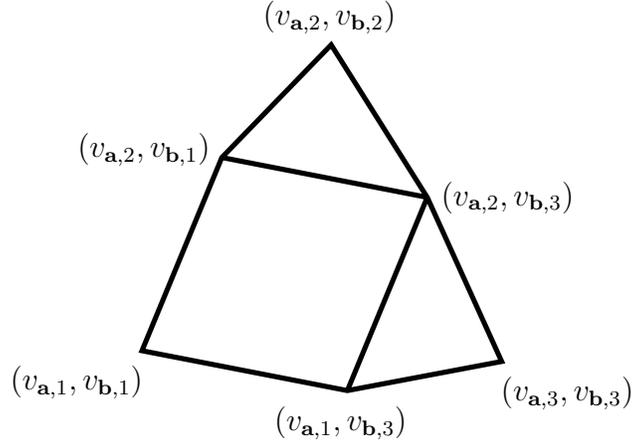
\begin{figure}
	\centering
	\begin{tikzpicture}
		\node[anchor=south west,inner sep=0] (image) at (0,0) {\includegraphics[width=.3\textwidth]{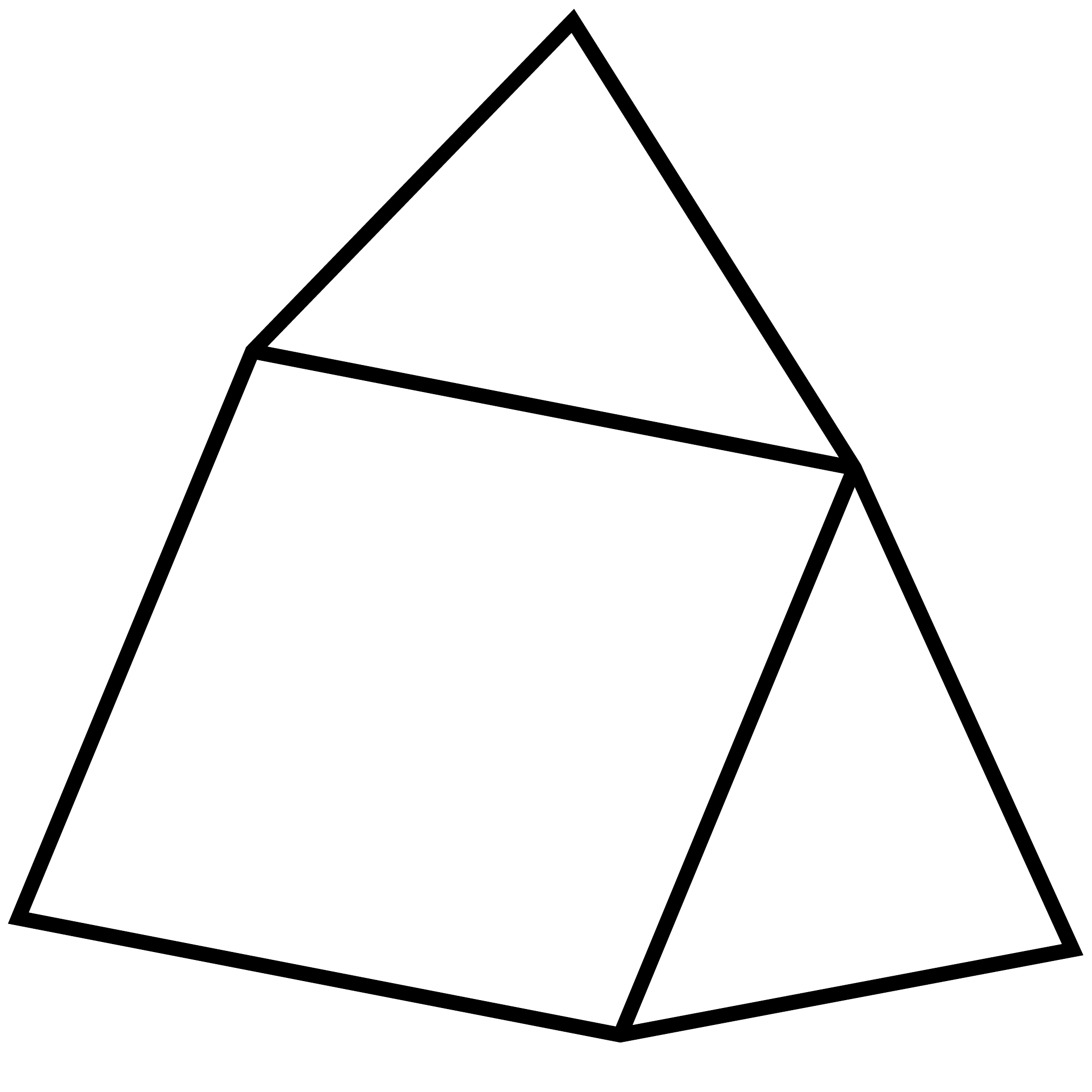}};
		\begin{scope}[shift={(image.south west)}, x={(image.south east)}, y={(image.north west)}]
			\node[anchor=north east] at (0.05, 0.15) {$(v_{{\bf a}, 1}, v_{{\bf b}, 1})$};
			\node[anchor=north] at (0.55, 0.04) {$(v_{{\bf a}, 1}, v_{{\bf b}, 3})$};
			\node[anchor=north west] at (0.95, 0.12) {$(v_{{\bf a}, 3}, v_{{\bf b}, 3})$};
			\node[anchor=west] at (0.79, 0.57) {$(v_{{\bf a}, 2}, v_{{\bf b}, 3})$};
			\node[anchor=south] at (0.52,0.98) {$(v_{{\bf a}, 2}, v_{{\bf b}, 2})$};
			\node[anchor=east] at (0.23,0.7) {$(v_{{\bf a}, 2}, v_{{\bf b}, 1})$};
		\end{scope}
	\end{tikzpicture}
	\caption{An example of a construction of a homotopy between $\Gamma_{\A_1\A_2} \triangleright \Gamma_{\A_2\A_3}$ and $\Gamma_{\A_1\A_3}$ for a specific choice of $\Gamma_{xy}$ using a square move and two triangle moves. }
\end{figure}

To construct the interpolation map associated with (\ref{gammathree}), which is composed of a sequence of elementary homotopy moves, we first choose a set interpolation maps of the form (\ref{iibar}) associated with each edge of a path $\Gamma$ that appear in the homotopy sequence, and then construct the interpolation map associated with each merger or square move between the $\Gamma$'s, subject to compatibility conditions as follows.

We associate the merger move of (\ref{mergeraa}) with an interpolation map
\ie\label{segmerg}
{\cal I}^{\rm M}: \Delta^2\times {\cal D}_{\A_1\A_2\A_3}\times \mathbb{R}^{0|*d_o} \to \mathfrak{U}_{\A_1\A_2\A_3},
\fe
where ${\cal I}^{\rm M}(\vec s, t, \nu)$ is the SRS obtained by deforming the split SRS over $\Sigma_t$ with a set of $d_o+2$ superconformal gluing maps along boundaries of super discs that depend on the odd parameter $\nu^a$, $a=1,\cdots,d_o$. Among them, the $d_o-1$ gluing maps that involve $\nu^b$ for $b\not={\bf a}$ are those of the standard form (\ref{superdiscglue}), where the center of the super disc $z_b$ is the $b$-th PCO location as specified by the vertex in $\Delta_b$ of the segment appearing in (\ref{mergeraa}). The gluing maps that involve $\nu^{\bf a}$, on the other hand, are defined on the boundaries of three super discs $D_1, D_2, D_3$ centered at the PCO locations associated with $v_{{\bf a},1}$, $v_{{\bf a},2}$, $v_{{\bf a},3}$. Let $(w_i,\eta_i)$ be the super coordinates on $D_i$. The gluing maps are of the form
\ie
w_i = z - {\theta P_i(\vec s) \nu^{\bf a} \over z-z_{\bf a}^{(i)}}, ~~~~ \eta_i = z - {P_i(\vec s) \nu^{\bf a}\over z-z_{\bf a}^{(i)}},
\fe
for $i=1,2,3$, $\vec s\in \Delta^2$. Here $P_i$ are a set of three complex valued smooth functions on $\Delta^2$,
subject to the following conditions.

Let us label the three vertices of the domain $\Delta^2$ of ${\cal I}^{\rm M}$ as $\{1,2,3\}$, and the edges as $\{\overline{12},\overline{23},\overline{13}\}$. We demand that ${\cal I}^{\rm M}|_{\overline{ij}}$ agrees with the interpolation map ${\cal I}_{ij}$ of the form (\ref{iibar}) already chosen for the edge $\overline{v_{{\bf a},i}v_{{\bf a},j}}$ in $\Gamma$ or $\Gamma'$, up to rescaling $\nu^{\bf a}$ by a non-vanishing complex valued function $Q_i(s)$ on $[0,1]$. In other words, let $R_{ij}(s)$ and $\widetilde R_{ij}(s)$ be the complex valued functions appearing in the gluing maps (\ref{wetag}), (\ref{wetatil}) associated with the interpolation map ${\cal I}_{ij}$, we demand
\ie\label{pppq}
P_1|_{\overline{12}}(s) = (1-s) R_{12}(s) Q_3(s),~~~ P_2|_{\overline{12}}(s) = s \widetilde R_{12}(s) Q_3(s),~~~ P_3|_{\overline{12}}(s) =0,
\fe
and similar relations related by cyclic permutation of the vertices $\{1,2,3\}$. By virtue of the construction of ${\cal I}_{ij}$ as in section \ref{sec:vertemerge}, the compatibility condition (\ref{pppq}) already ensures that ${\cal I}^{\rm M}$ has non-singular Berezinian when restricted to $\partial \Delta^2$. However, we must ensure that this is the case when extending ${\cal I}^{\rm M}$ to the interior of its domain $\Delta^2$.

By a similar argument as below (\ref{sumgs}), the Berezinian of ${\cal I}^{\rm M}$ would be singular if there exists a weight $-{1\over 2}$ meromorphic differential $\zeta=f(z) dz^{-{1\over 2}}$ with $d_o+2$ poles at $\{z_b: 1\leq b\leq d_o, b\not={\bf a}\}\cup\{z_{\bf a}^{(i)}, i=1,2,3\}$, whose residue at $z_{\bf a}^{(i)}$ is equal to $P_i(\vec s)$ for some $\vec s\in\Delta^2$. This is equivalent to the locus
\ie\label{cpzero}
\sum_{i=1}^3 c_i P_i(\vec s) = 0
\fe
where $c_i$ are given by the inverse of the correlators $\big\langle\delta(\B(z_{\bf a}^{(i)}) \prod_{b\not={\bf a}} \delta(\B(z_b))\big\rangle$ of the $\B\C$ system. While we already know that $\sum_{i=1}^3 c_i P_i(\vec s)$ is non-vanishing on $\partial\Delta^2$, a priori the phase of the former could have a nonzero winding number around $\partial\Delta^2$ which would force (\ref{cpzero}) somewhere in the interior of $\Delta^2$. By choosing the functions $Q_i(s)$ appearing in (\ref{pppq}), however, such a winding number can be eliminated, which allows for extending $P_i(\vec s)$ to $\Delta^2$ while evading (\ref{cpzero}).

Having constructed the interpolation map for the merger move, a calculation similar to section \ref{sec:vertemerge} shows that it does not contribute to the integral of $\Omega$. We now turn to the interpolation map of the square move (\ref{squareab}), of the form
\ie{}
{\cal I}^{\rm Sq}: [0,1]^2\times {\cal D}_{\A_1 \A_2 \A_3}\times \mathbb{R}^{0|*d_o} \to \mathfrak{U}_{\A_1 \A_2 \A_3}.
\fe
${\cal I}^{\rm Sq}(\vec s,t,\nu)$ is the SRS defined by deforming gluing maps on the boundary of $d_o+2$ discs. Among them, $d_o-2$ gluing maps that involve $\nu^c$ for $c\not={\bf a}, {\bf b}$ are on the boundary of discs centered at $z_c$ as specified by the vertex in $\Delta_c$ of the segment of the paths appearing in (\ref{squareab}). The gluing maps that involve $\nu^{\bf a}$ and $\nu^{\bf b}$ are defined on the boundaries of four super discs $D_{ij}$, $i,j=1,2$, centered at PCO locations $\{z_{\bf a}^{(2)}, z_{\bf a}^{(3)},z_{\bf b}^{(1)},z_{\bf b}{}^{(2)} \}\equiv \{z_{11},z_{12},z_{21},z_{22}\}$ associated with the vertices $v_{{\bf a},2}$, $v_{{\bf a},3}$, $v_{{\bf b} ,1}$, $v_{{\bf b},2}$. Let $(w_{ij}, \eta_{ij})$ be the supercoordinates on $D_{ij}$. The gluing maps are of the form
\ie
w_{ij} = z - {\theta  P_{ij}(\vec s)\cdot \nu\over z-z_{ij}},~~~~ \eta_{ij} = z-{ P_{ij}(\vec s)\cdot\nu\over z-z_{ij}}
\fe
for $i,j=1,2$, where we used the notation $\nu=(\nu^{\bf a},\nu^{\bf b})$. Each $P_{ij}(\vec s)\equiv (P_{ij1}(\vec s), P_{ij2}(\vec s))$ is a $\mathbb{C}^2$-valued function on $[0,1]^2$, subject to the boundary conditions
\ie\label{fourqs}
& P_{ij\ell} (0,s) = \begin{pmatrix} R_{11}(0)  & 0 \\ (1-s) R_{21}(s)  & s \widetilde R_{21}(s)  \end{pmatrix}_{ij} (Q_1)_{i\ell}(s),
~~~P_{ij\ell}(1,s) = \begin{pmatrix} 0 & \widetilde R_{11}(1) \\ (1-s) R_{22}(s) & s \widetilde R_{22}(s) \end{pmatrix}_{ij}(Q_3)_{i\ell}(s) ,
\\
& P_{ij\ell} (s,0) = \begin{pmatrix}  (1-s) R_{11}(s)  & s \widetilde R_{11}(s) \\ R_{21}(0) & 0 \end{pmatrix}_{ij} (Q_2)_{i\ell}(s),
~~~ P_{ij\ell} (s,1) = \begin{pmatrix} (1-s) R_{12}(s)  & s \widetilde R_{12}(s) \\ 0 & \widetilde R_{21}(1)  \end{pmatrix}_{ij} (Q_4)_{i\ell}(s),
\fe
where the functions $R_{ij},\widetilde R_{ij}$ for $ij=\{11,21\}$ and $\{12,22\}$ are those appearing in the gluing maps (\ref{wetag}), (\ref{wetatil}) used in constructing the interpolation map attached to the edges $\{\overline{v_{{\bf a},2}v_{{\bf a},3}}, \overline{v_{{\bf b},1}v_{{\bf b},2}}\}$ in $\Gamma$ and $\Gamma'$, respectively. $Q_1,\cdots,Q_4$ are $GL(2,\mathbb{C})$-valued functions on $[0,1]$ that are chosen to allow for extending $P_{ij\ell}$ smoothly into the interior of $[0,1]^2$ while ensuring that ${\cal I}^{\rm Sq}$ has non-singular Berezinian.

\bigskip

\begin{figure}[h]
	\centering
	\begin{tikzpicture}
		\node[anchor=south west,inner sep=0] (image) at (0,0) {\includegraphics[width=.7\textwidth]{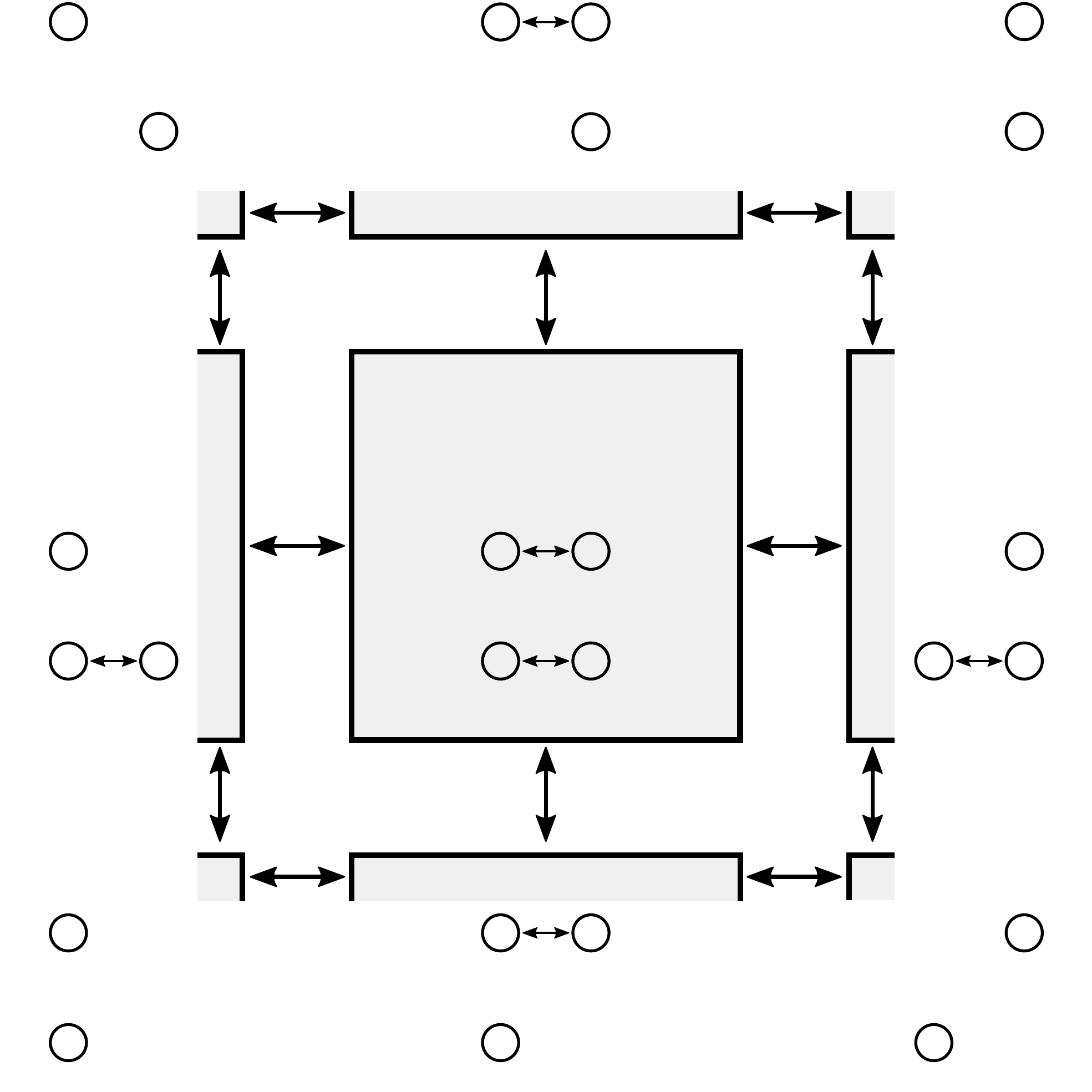}};
		\begin{scope}[shift={(image.south west)}, x={(image.south east)}, y={(image.north west)}]
			\node[anchor=north] at (0.000 + 0.063, 0.000 + 0.130) {$\scriptstyle 11$};
			\node[anchor=south] at (0.000 + 0.063, 0.000 + 0.057) {$\scriptstyle 21$};

			\node[anchor=north] at (0.396 + 0.063, 0.000 + 0.130) {$\scriptstyle 11$};
			\node[anchor=north] at (0.396 + 0.145, 0.000 + 0.130) {$\scriptstyle 12$};
			\node[anchor=south] at (0.396 + 0.063, 0.000 + 0.057) {$\scriptstyle 21$};
			\node[anchor=north] at (0.792 + 0.145, 0.000 + 0.130) {$\scriptstyle 12$};
			\node[anchor=south] at (0.792 + 0.063, 0.000 + 0.057) {$\scriptstyle 21$};

			\node[anchor=north] at (0.000 + 0.063, 0.350 + 0.130) {$\scriptstyle 11$};
			\node[anchor=south] at (0.000 + 0.063, 0.350 + 0.057) {$\scriptstyle 21$};
			\node[anchor=south] at (0.000 + 0.145, 0.350 + 0.057) {$\scriptstyle 22$};
			
			\node[anchor=north] at (0.396 + 0.063, 0.350 + 0.130) {$\scriptstyle 11$};
			\node[anchor=north] at (0.396 + 0.145, 0.350 + 0.130) {$\scriptstyle 12$};
			\node[anchor=south] at (0.396 + 0.063, 0.350 + 0.057) {$\scriptstyle 21$};
			\node[anchor=south] at (0.396 + 0.145, 0.350 + 0.057) {$\scriptstyle 22$};

			\node[anchor=north] at (0.792 + 0.145, 0.350 + 0.130) {$\scriptstyle 12$};
			\node[anchor=south] at (0.792 + 0.063, 0.350 + 0.057) {$\scriptstyle 21$};
			\node[anchor=south] at (0.792 + 0.145, 0.350 + 0.057) {$\scriptstyle 22$};
			
			\node[anchor=north] at (0.000 + 0.063, 0.834 + 0.130) {$\scriptstyle 11$};
			\node[anchor=south] at (0.000 + 0.145, 0.834 + 0.057) {$\scriptstyle 22$};
			
			\node[anchor=north] at (0.396 + 0.063, 0.834 + 0.130) {$\scriptstyle 11$};
			\node[anchor=north] at (0.396 + 0.145, 0.834 + 0.130) {$\scriptstyle 12$};
			\node[anchor=south] at (0.396 + 0.145, 0.834 + 0.057) {$\scriptstyle 22$};

			\node[anchor=north] at (0.792 + 0.145, 0.834 + 0.130) {$\scriptstyle 12$};
			\node[anchor=south] at (0.792 + 0.145, 0.834 + 0.057) {$\scriptstyle 22$};
			
			\node[anchor=south] at (0.5, 0.53) {\large $\displaystyle \mathcal{I}^{\rm Sq}$};
			\node[anchor=south] at (0.5, 0.32) {\small $s_2 = 0$};
			\node[anchor=north] at (0.5, 0.68) {\small $s_2 = 1$};
			
			\node[anchor=south, rotate=-90] at (0.32, 0.5) {\small $s_1 = 0$};
			\node[anchor=north, rotate=-90] at (0.68, 0.5) {\small $s_1 = 1$};
			
			\node[anchor=south] at (0.272, 0.5) {$Q_1$};
			\node[anchor=south] at (0.728, 0.5) {$Q_3$};
			\node[anchor=east] at (0.5, 0.272) {$Q_2$};
			\node[anchor=east] at (0.5, 0.728) {$Q_4$};
		\end{scope}
	\end{tikzpicture}
	\caption{An illustration of the interpolation map ${\cal I}^{\rm Sq}$ associated with a square move. The shaded square at the center represents the domain of the interpolation parameters $(s_1,s_2)$. The neighboring shaded domains are those of other interpolation maps. In each region, the (up to four) circles represent the nontrivial gluing maps of the discs centered at $z_{ij}$ on the SRS. An arrow pointing between two circles indicates interpolation between the gluing maps as $s_1$ or $s_2$ varies, in accordance with the pattern of PCO movement. Matching the boundaries segments of ${\cal I}^{\rm Sq}$ to the neighoring segments of the integration contour involves further $GL(2,\mathbb{C})$ rotations on the fermionic coordinates $(\nu^{\bf a}, \nu^{\bf b})$ by $Q_i$.}
\end{figure}
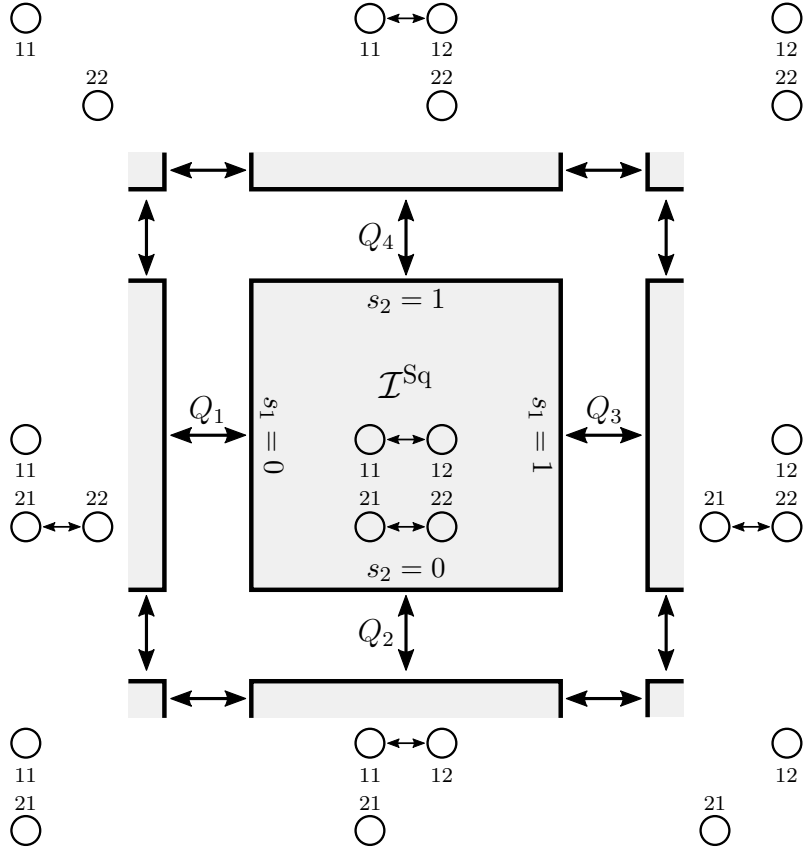

More explicitly, the Berezinian of ${\cal I}^{\rm Sq}$ would be singular if the $2\times 2$ matrix
\ie\label{ijps}
A^m{}_\ell(\vec s) = \sum_{i,j=1}^2 c_{ij}^m P_{ij\ell}(\vec s) 
\fe
is singular, where the coefficients $c_{ij}^m$ are such that $\sum_{i,j} c_{ij}^m f^{(-1)}(z_{ij})=0$ for all weight $-{1\over 2}$ meromorphic differentials $\zeta = f(z) dz^{-{1\over 2}}$ with $d_o+2$ poles at $\{z_c: 1\leq c\leq d_o, c\not={\bf a},{\bf b}\} \cup \{z_{ij}: i,j=1,2\}$. Up to overall rescaling, such $c^m_{ij}$ are related to $\B\C$ system correlators via
\ie 
c_{ij}^1 c_{i'j'}^2 - c^2_{ij} c^1_{i'j'} = \Big\langle\delta(\B(z_{ij})) \delta(\B(z_{i'j'})) \prod_{c\not={\bf a},{\bf b}} \delta(\B(z_c))\Big\rangle^{-1}.
\fe
We can choose $Q_1,\cdots, Q_4$ in (\ref{fourqs}) to eliminate the winding number of $A^m{}_{\ell}(\vec s)$ in $GL(2,\mathbb{C})$ as $\vec s$ moves around the boundary of the square $[0,1]^2$. This then removes the obstruction in extending $P_{ij\ell}$ to $[0,1]^2$ while maintaining the non-degeneracy of (\ref{ijps}).

The integration of $({\cal I}^{\rm Sq})^*\Omega$ with respect to $\vec s\in [0,1]^2$ and $\nu^{\bf a}, \nu^{\bf b}$ can be performed analogously to the computation of section \ref{sec:vertemerge}, and produces the factor
\ie
\Big[ \xi(z_{\bf a}^{(3)}) - \xi(z_{\bf a}^{(2)}) \Big] \Big[ \xi(z_{\bf b}^{(2)}) - \xi(z_{\bf b}^{(1)}) \Big]
\fe
in precise agreement with the vertical integration prescription associated with moving a pair of holomorphic PCOs, over a codimension 2 locus in the bosonic moduli space.

The above construction can be generalized inductively to $\mathfrak{I}_{\A_1\A_2\cdots \A_{p+1}}$, which is associated with a ``$p$-path"
\ie\label{gammapis}
\Gamma_{\A_1\A_2\cdots \A_{p+1}}: \Delta^p \to V,
\fe
whose restriction to the $m$-th face $\partial_m\Delta^p$ agrees with $\Gamma_{\A_1\cdots \A_{m-1} i_{m+1}\cdots \A_{p+1}}$, $m=1,\cdots,p+1$. Viewed as a homotopy, (\ref{gammapis}) is composed of merger moves that do not contribute to the supermoduli integral, and the $p$-dimensional analog of square moves, or ``hypercube moves". Each hypercube move is associated with an embedded hypercube of the form 
\ie
\prod_{m=1}^p\overline{v_{{\bf a}_m, p+1-m}v_{{\bf a}_m, p+2-m}} \,\subset \prod_{m=1}^p \Delta_{{\bf a}_m},
\fe
and gives rise to an interpolation map
\ie
{\cal I}^{\rm HC} : [0,1]^p\times {\cal D}_{\A_1\cdots\A_{p+1}} \times \mathbb{R}^{0|*d_o}
\to \mathfrak{U}_{\A_1\cdots\A_{p+1}},
\fe
such that the integral of $({\cal I}^{\rm HC})^*\Omega$ with respect to $\vec s\in [0,1]^p$ and the odd parameters $\nu^{{\bf a}_1},\cdots, \nu^{{\bf a}_p}$ gives rise to the factor
\ie
\prod_{m=1}^p \Big[ \xi(z_{{\bf a}_m}^{(p+2-m)}) -  \xi(z_{{\bf a}_m}^{(p+1-m)}) \Big]
\fe
that amounts to the vertical integration associated with moving $p$ holomorphic PCOs.

\section{Concluding remarks}
\label{sec:discuss}

While the emergence of PCO from integrating over the fermionic moduli parameter in the gluing map of the super disc is well known (see e.g. \cite{Polchinski:1998rr}), a key point of the construction of section \ref{sec:mtos} is that the interpolation map $\mathfrak{I}_{\A\B}$ that defines the segment of the supermoduli integration contour at the interface between the cells ${\cal D}_\A$ and ${\cal D}_\B$ involves not actual continuous movement of the PCO location, but rather deforming simultaneously the gluing maps on a pair of discs corresponding to the initial and final positions of the PCO in the vertical integration. Furthermore, it is important that $\mathfrak{I}_{\A\B}$ need not serve as interpolation, in the sense of a continuous map, between the coordinate maps $\varphi_\A$ and $\varphi_\B$. Rather, we only need to demand that the boundary of the image of $\mathfrak{I}_{\A\B}$ to cancel against the boundaries of ${\cal D}_\A$ and ${\cal D}_\B$ lifted into the supermoduli space, so as to close the supermoduli contour $\mathfrak{S}$ into a cycle. This relaxation in the interpolation condition, characterized by the freedom of the super-diffeomorphism $R$ appearing in (\ref{intmatchcond}) or (\ref{supcontr}), allows for the construction of the interpolation maps over arbitrary codimension interfaces with non-singular Berezenian without obstruction.

In a follow-up paper \cite{paper:g2}, we will illustrate the general construction of section \ref{sec:mtos} in the example of genus two SRSs, for both odd and even spin structures, with explicit parameterization of the supermoduli space related to specific PCO configurations.

Given the understanding of this paper, one may hope to recast closed superstring field theory, currently formulated in the PCO language \cite{deLacroix:2017lif}, in terms of SRS and supermoduli integration. Another possible future direction is to find a local projection of the supermoduli space that recovers the bosonic moduli space integrand of the pure spinor formalism \cite{Berkovits:2000fe, Berkovits:2002zk, Berkovits:2013eqa}.

\section*{Acknowledgements}

XY thanks Ashoke Sen for conversations that motivated this work. We would like to thank Ashoke Sen for reading a preliminary draft, and Ted Erler for correspondence.
This work is supported in part by a Simons Investigator Award from the Simons Foundation, by the Simons Collaboration Grant on the Non-Perturbative Bootstrap, and by DOE grants DE-SC0007870.

\appendix

\section{Explicit verification of the interpolation integral identity}
\label{sec:verifyint}

Here we explicitly verify (\ref{onedimbreak}) with the following choice of $\mathfrak{I}$,
\ie
\mathfrak{I}(s,\nu) = \varphi(s\, r(\nu), \ell_s(\nu)), 
\fe
where $\ell_s$ is a 1-parameter family of $GL(n_o,\mathbb{C})$ rotations on the $\nu^a$'s, and $r(\nu)$ is a Grassmann-even function of $\nu^a$ that obeys
\ie\label{varphishift}
r(\nu) + f(r(\nu), \ell_1( \nu)) = 0,
\fe
so that 
\ie
\varphi^{-1}\circ \mathfrak{I}(0,\nu) = (0,\ell_0(\nu)),~~~~ \widetilde\varphi^{-1}\circ {\mathfrak I}(1,\nu) = (0, g(r(\nu), \ell_1(\nu)) ).
\fe
By assumption, $R(\nu) = \ell_0(\nu)$ and $\widetilde R(\nu) = g(r(\nu),\ell_1(\nu))$ are super-diffeomorphisms. Note that $r(\nu)$ is determined by expanding (\ref{varphishift}) order by order in $\nu^a$. Explicitly, for the  form
\ie
\varphi^*\Omega = \omega(t,\nu) [dt | d^{n_o} \nu],
\fe
the interpolation term in (\ref{onedimbreak}) can be written, using the Berezinian of $\varphi^{-1}\circ \mathfrak{I}$, as
\ie\label{intintsimp}
& \int_{ [0,1]\times \mathbb{R}^{0|n_o}} \mathfrak{I}^*\Omega 
\\
& = \int_0^1 [ds |d^{n_o} \nu]\,\left[ r(\nu) + s {\partial r(\nu)\over\partial\nu^b} ((\partial_\nu \ell_s)^{-1})^b{}_a \partial_s \ell_s^a(\nu) \right] \left( \det {\partial\ell_s^a(\nu)\over \partial \nu^b} \right)^{-1} \omega(s r(\nu), \ell_s(\nu)).
\fe

Let us consider the simplest nontrivial case $n_o=2$, with the transition map (\ref{transitionmaptf}) given by
\ie
\widetilde t = t + f_{12}(t) \nu^1\nu^2, ~~~~\widetilde \nu^a=\nu^a.
\fe
We will take $\ell_s^a(\nu) = R^a{}_b(s) \nu^b$, and solve (\ref{varphishift}) with $r(\nu) = - f_{12}(0) \det(R(1))\nu^1 \nu^2$. The interpolation integral (\ref{intintsimp}) evaluates to
\ie\label{simpintin}
& \int_0^1 [ds | d^2\nu] \left[  - f_{12}(0) \det(R(1))\nu^1 \nu^2 \right] \left[ 1 - s\, {\rm tr}(R(s)^{-1}\partial_s R(s)) \right](\det R(s))^{-1} \omega(0,0) 
\\
& = - f_{12}(0) \omega(0,0).
\fe
The first two terms on the RHS of (\ref{onedimbreak}) are
\ie{}
& \int_{t<0} [dt|d^2\nu] \omega(t,\nu) + \int_{\widetilde t>0} [d\widetilde t |d^2\nu] (1+ f'_{12}(\widetilde t) \nu^1 \nu^2)^{-1}  \omega(\widetilde t-f_{12}(\widetilde t) \nu^1 \nu^2,\nu)
\\
& = \int_{-\infty}^\infty [dt|d^2\nu] \omega(t,\nu)  - \int_{\widetilde t>0} d\widetilde t \left[ f_{12}(\widetilde t) \partial_t \omega(\widetilde t ,0) + f_{12}'(\widetilde t)\omega(\widetilde t,0) \right] 
\fe
The last terms on the RHS cancels against (\ref{simpintin}), leaving the integral $\int [dt |d^2\nu] \omega(t,\nu)$ as desired.

\bibliographystyle{JHEP}
\bibliography{srs}

\providecommand{\href}[2]{#2}\begingroup\raggedright\begin{thebibliography}{10}

\bibitem{DHoker:1988pdl}
E.~D'Hoker and D.~H. Phong, {\it {The Geometry of String Perturbation Theory}},
   {\em Rev. Mod. Phys.} {\bf 60} (1988) 917.

\bibitem{Witten:2012ga}
E.~Witten, {\it {Notes On Super Riemann Surfaces And Their Moduli}},  {\em Pure
  Appl. Math. Quart.} {\bf 15} (2019), no.~1 57--211,
  [\href{http://arxiv.org/abs/1209.2459}{{\tt arXiv:1209.2459}}].

\bibitem{Friedan:1985ge}
D.~Friedan, E.~J. Martinec, and S.~H. Shenker, {\it {Conformal Invariance,
  Supersymmetry and String Theory}},  {\em Nucl. Phys. B} {\bf 271} (1986)
  93--165.

\bibitem{Verlinde:1987sd}
E.~P. Verlinde and H.~L. Verlinde, {\it {Multiloop Calculations in Covariant
  Superstring Theory}},  {\em Phys. Lett. B} {\bf 192} (1987) 95--102.

\bibitem{Atick:1987rk}
J.~J. Atick, G.~W. Moore, and A.~Sen, {\it {Some Global Issues in String
  Perturbation Theory}},  {\em Nucl. Phys. B} {\bf 308} (1988) 1--101.

\bibitem{Sen:2014pia}
A.~Sen, {\it {Off-shell Amplitudes in Superstring Theory}},  {\em Fortsch.
  Phys.} {\bf 63} (2015) 149--188, [\href{http://arxiv.org/abs/1408.0571}{{\tt
  arXiv:1408.0571}}].

\bibitem{Sen:2015hia}
A.~Sen and E.~Witten, {\it {Filling the gaps with PCO\textquoteright{}s}},
  {\em JHEP} {\bf 09} (2015) 004, [\href{http://arxiv.org/abs/1504.00609}{{\tt
  arXiv:1504.00609}}].

\bibitem{Witten:2012bg}
E.~Witten, {\it {Notes On Supermanifolds and Integration}},  {\em Pure Appl.
  Math. Quart.} {\bf 15} (2019), no.~1 3--56,
  [\href{http://arxiv.org/abs/1209.2199}{{\tt arXiv:1209.2199}}].

\bibitem{Donagi:2013dua}
R.~Donagi and E.~Witten, {\it {Supermoduli Space Is Not Projected}},  {\em
  Proc. Symp. Pure Math.} {\bf 90} (2015) 19--72,
  [\href{http://arxiv.org/abs/1304.7798}{{\tt arXiv:1304.7798}}].

\bibitem{deLacroix:2017lif}
C.~de~Lacroix, H.~Erbin, S.~P. Kashyap, A.~Sen, and M.~Verma, {\it {Closed
  Superstring Field Theory and its Applications}},  {\em Int. J. Mod. Phys. A}
  {\bf 32} (2017), no.~28n29 1730021,
  [\href{http://arxiv.org/abs/1703.06410}{{\tt arXiv:1703.06410}}].

\bibitem{Erler:2017dgr}
T.~Erler and S.~Konopka, {\it {Vertical Integration from the Large Hilbert
  Space}},  {\em JHEP} {\bf 12} (2017) 112,
  [\href{http://arxiv.org/abs/1710.07232}{{\tt arXiv:1710.07232}}].

\bibitem{paper:g2}
C.~Wang and X.~Yin, {\it {Supermoduli Space and PCOs at Genus Two}},  {\em to
  appear}.

\bibitem{Polchinski:1998rr}
J.~Polchinski, {\em {String theory. Vol. 2: Superstring theory and beyond}}.
\newblock Cambridge Monographs on Mathematical Physics. Cambridge University
  Press, 12, 2007.

\bibitem{Berkovits:2000fe}
N.~Berkovits, {\it {Super Poincare covariant quantization of the superstring}},
   {\em JHEP} {\bf 04} (2000) 018,
  [\href{http://arxiv.org/abs/hep-th/0001035}{{\tt hep-th/0001035}}].

\bibitem{Berkovits:2002zk}
N.~Berkovits, {\it {ICTP lectures on covariant quantization of the
  superstring}},  {\em ICTP Lect. Notes Ser.} {\bf 13} (2003) 57--107,
  [\href{http://arxiv.org/abs/hep-th/0209059}{{\tt hep-th/0209059}}].

\bibitem{Berkovits:2013eqa}
N.~Berkovits, {\it {Covariant Map Between Ramond-Neveu-Schwarz and Pure Spinor
  Formalisms for the Superstring}},  {\em JHEP} {\bf 04} (2014) 024,
  [\href{http://arxiv.org/abs/1312.0845}{{\tt arXiv:1312.0845}}].

\end{thebibliography}\endgroup

\end{document}